\documentclass[structabstract,letter]{aa} 
\usepackage{graphicx}
\usepackage{float}
\usepackage{txfonts}
\usepackage{natbib}
\usepackage{graphicx} 
\usepackage{float}
\usepackage{txfonts}
\usepackage{natbib}
\usepackage{units}
\usepackage{url}

\newcommand{\xte}{XTE J1752-223}
\newcommand{\lsp}{LS~I~+61$^{\circ}$303}
\newcommand{\lsi}{LS~I~+61$^{\circ}$303~}

\newcommand{\psr}{PSR B1259$-$63}
\newcommand{\grs}{GRS 1915+105~}
\newcommand{\beq}{\begin{equation}}
\newcommand{\eneq}{\end{equation}}

\begin{document}

\title{The broad-band radio spectrum of \lsi in outburst} 

\author{L. Zimmermann\inst{}, L. Fuhrmann\thanks{corresponding author: lfuhrmann@mpifr-bonn.mpg.de}\inst{}\and M. Massi\inst{}}

\institute{Max Planck Institut f\"ur Radioastronomie, Auf dem H\"ugel 69, 53121 Bonn, Germany}

\date{Received 2014; }

\abstract 
{} 
{Our aim is to explore the broad-band radio continuum spectrum of
  \lsi{} during its outbursts by employing the available set of
  secondary focus receivers of the Effelsberg 100 m telescope.}
{The clear periodicity of the system \lsi{} allowed observations to be
  scheduled covering the large radio outburst in March-April 2012. We
  observed \lsi on 14 consecutive days at 2.6, 4.85, 8.35, 10.45,
  14.3, 23, and 32 GHz with a cadence of about 12 hours followed by
  two additional observations several days later. Based on these
  observations we obtained a total of 24 quasi-simultaneous broad-band
  radio spectra.}
{During onset, the main flare shows an almost flat broad-band
  spectrum, most prominently seen on March 27, 2012, where -- for the
  first time -- a flat spectrum ($\alpha=0.00\pm0.07$,
  $S\propto\nu^{\alpha}$) is observed up to 32\,GHz (9\,mm
  wavelength). The flare decay phase shows superimposed ``sub-flares''
  with the spectral index oscillating between $-$0.4 and $-$0.1 in a
  quasi-regular fashion. Finally, the spectral index steepens during
  the decay phase, showing optically thin emission with values
  $\alpha\sim$\,$-$0.5 to $-$0.7.}
{The radio characteristics of \lsi compare well with those of the
  microquasars XTE J1752-223 and Cygnus X-3. In these systems the
  flaring phase is actually also composed of a sequence of
  outbursts with clearly different spectral characteristics: a first
  outburst with a flat/inverted spectrum followed by a bursting phase
  of optically thin emission.}


\titlerunning{The broad-band radio spectrum of \lsi in outburst}
\maketitle

\section{Introduction}

The periodical radio emitting stellar system \lsi consists of a fast
rotating Be star and a compact object of unclear nature.  One
hypothesis is that the compact object is a radio pulsar and that the
relativistic electrons responsible for the radio outburst are produced
through an interaction between the relativistic wind of the pulsar and
the wind of the Be star \citep{maraschitreves81, dhawan06, dubus06},
i.e. similar to the system \psr{}. This system is formed by a O9.5Ve
star and a compact object with well-detected radio pulses, i.e. a
radio pulsar. The radio outburst of \psr{} occurring around periastron
passage is explained by electrons accelerated at the shock front
between the two winds \citep[][and references therein]{chernyakova14}.
For the compact object in \lsi with the radio outburst towards
apastron, however, an alternative hypothesis suggests an accreting
object, i.e. a black hole or a neutron star
\citep[e.g.][]{taylor92,romero07,massikaufman09}. In this case, the
radio emitting electrons belong to a jet typical of microquasars. The
\citet{bondihoyle44} accretion theory for the eccentric orbit of
\lsi{} with an orbital period $P=$26.496\,days (phase $\Phi$) then
predicts two accretion maxima and therefore two ejections of
relativistic electrons \citep{taylor92, martiparedes95,
  boschramon06,romero07}.  During the first ejection near periastron
\citep[$\Phi_{\rm peri}=0.23 \pm 0.02$,][]{casares05} relativistic
electrons suffer severe inverse-Compton (IC) losses due to the
proximity to the Be star. Consequently, a high energy outburst is
expected, as indeed confirmed by {\it Fermi}/LAT observations
\citep{abdo09}, but no or only negligible radio emission
\citep{boschramon06}.  For the second accretion peak near apastron it
has been demonstrated by \citet{boschramon06} that the larger distance
between the compact object and the Be star results in lower IC losses
(i.e. a smaller high energy outburst) and the relativistic electrons
can propagate out of the orbital plane into a radio jet. Recent {\it
  Fermi}/LAT observations also confirmed this second, smaller high
energy outburst \citep{jaronmassi14}.

\begin{figure*}[t]
\centering
\includegraphics[width=9.7cm,angle=0]{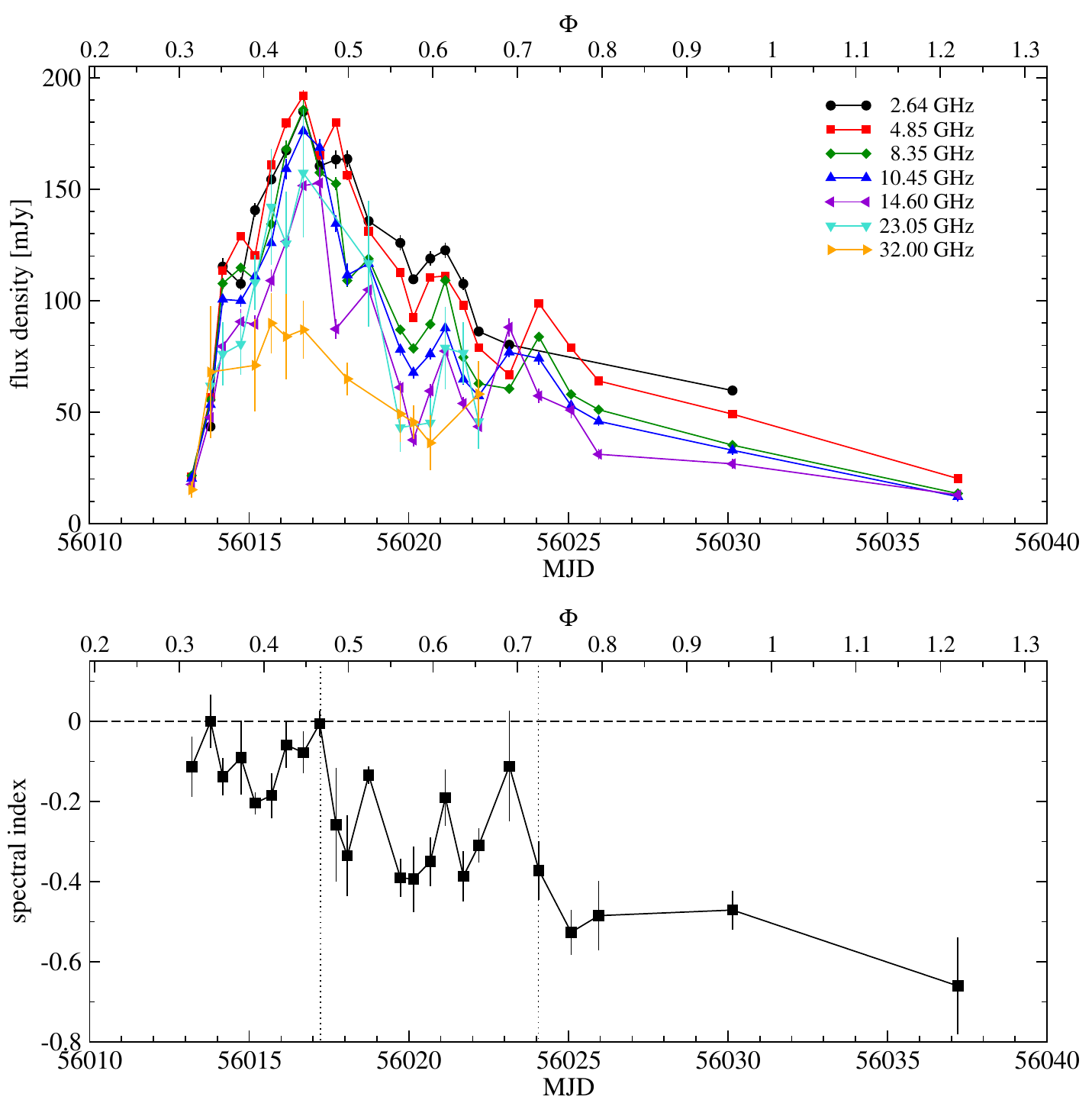}
\caption{Top: Radio outburst of \lsp: flux density (mJy) vs.  time
  (MJD) and orbital phase $\Phi$ as observed with the Effelsberg 100 m
  telescope at seven frequencies between 2.6\,GHz and 32\,GHz.  Bottom:
  Broad-band spectral index $\alpha$ (all frequencies).}
\label{LCs}
\end{figure*}

Synchrotron radiation emitted from one relativistic electron
population with density $N_{\rm rel}$ gyrating along a magnetic field
$B$ produces a power-law radio spectrum ($S\propto\nu^{\alpha}$) with
spectral slope $\alpha_{\rm thin} < 0$ above a critical frequency
$\nu_{break}$ that depends on $B$ and $N_{\rm rel}$.  Below this
critical frequency self-absorption effects become important and the
emission becomes optically thick with a spectral slope $\alpha_{\rm
  thick}=2.5$ \citep[e.g.][]{kaiser06}.  For instance, the spectrum of
the radio outburst in the pulsar system \psr{} has a spectral index
$\alpha=-0.7$, which is consistent with optically thin synchrotron
emission \citep[Fig.~3 in][]{connors02}.  In contrast, the jets of
microquasars often show flat/inverted radio spectra as demonstrated by
\citet{fender01}. Plasma and magnetic field variations along the jet
in fact may create regions with different $B$ and $N_{\rm rel}$ values
resulting in spectral components with different turn-over frequencies
$\nu_{break}$. The overlap of the optically thin part of one spectral
component with the self-absorbed part of the adjacent one will result
in a flat spectrum when observed with a spatial resolution
insufficient to resolve the jet \citep{kaiser06}.  In the microquasar
Cygnus X-1, radio emission has been measured with a flat spectrum,
i.e.  $\alpha=-0.06 \pm 0.05$ and $\alpha=0.07 \pm 0.04$
\citep{fender00}. In Cygnus X-1 the black hole is powered by accretion
of the stellar wind of its supergiant companion star. Since the
companion is close to filling its Roche lobe, the wind is not
symmetric but strongly focused towards the black hole \citep[and
references therein]{miskovica11}.  In contrast, as discussed above in
the context of an accretion scenario, accretion in \lsi{} occurs in
two particular orbital phases due to the eccentric orbit around the Be
star.  Is the accretion phase and subsequent ejection in \lsi{} able
to develop a microquasar jet with a flat spectrum? As shown by
\citet{corbel13}, in the early phase of jet formation the low density
particles in the jet produce an optically thin synchrotron power-law
spectrum at GHz frequencies. Only when the density of the jet plasma
increases, a transition to higher optical depth occurs resulting in
flat/inverted radio spectra ($\alpha \geq 0$). In \lsi strong evidence
for a flat radio spectrum does exist. Early VLA observations at four
epochs revealed a flat spectrum between 1.5 and 22\,GHz
\citep{gregory79}. Furthermore, \citet[][]{1998ApJ...497..419S}
carried out multi-frequency VLA observations sparsely covering one
orbit and found deviations in the radio spectrum from a simple power
law during the outburst. Finally, \citet{massikaufman09} measured even
inverted, optically thick spectra between 2.2 GHz and 8.3 GHz.  The
primary aim of the current work is to extend previous radio
observations of \lsi and systematically study -- for the first time --
broad-band cm/mm (2.6--32\,GHz) radio spectra and their evolution
during a complete outburst. The densely sampled observations of \lsi
were performed over a period of about three weeks (a total of
24\,days) using the Effelsberg 100 m telescope at a total of seven
frequency bands.  The paper is structured as follows. In Sect. 2 we
briefly present the observations and data reduction procedures. In
Sect. 3 we present the results.  Section 4 provides a short discussion
and our conclusions.
 
\begin{figure*}[ht]
\center
\includegraphics[width=3.4cm,angle=0]{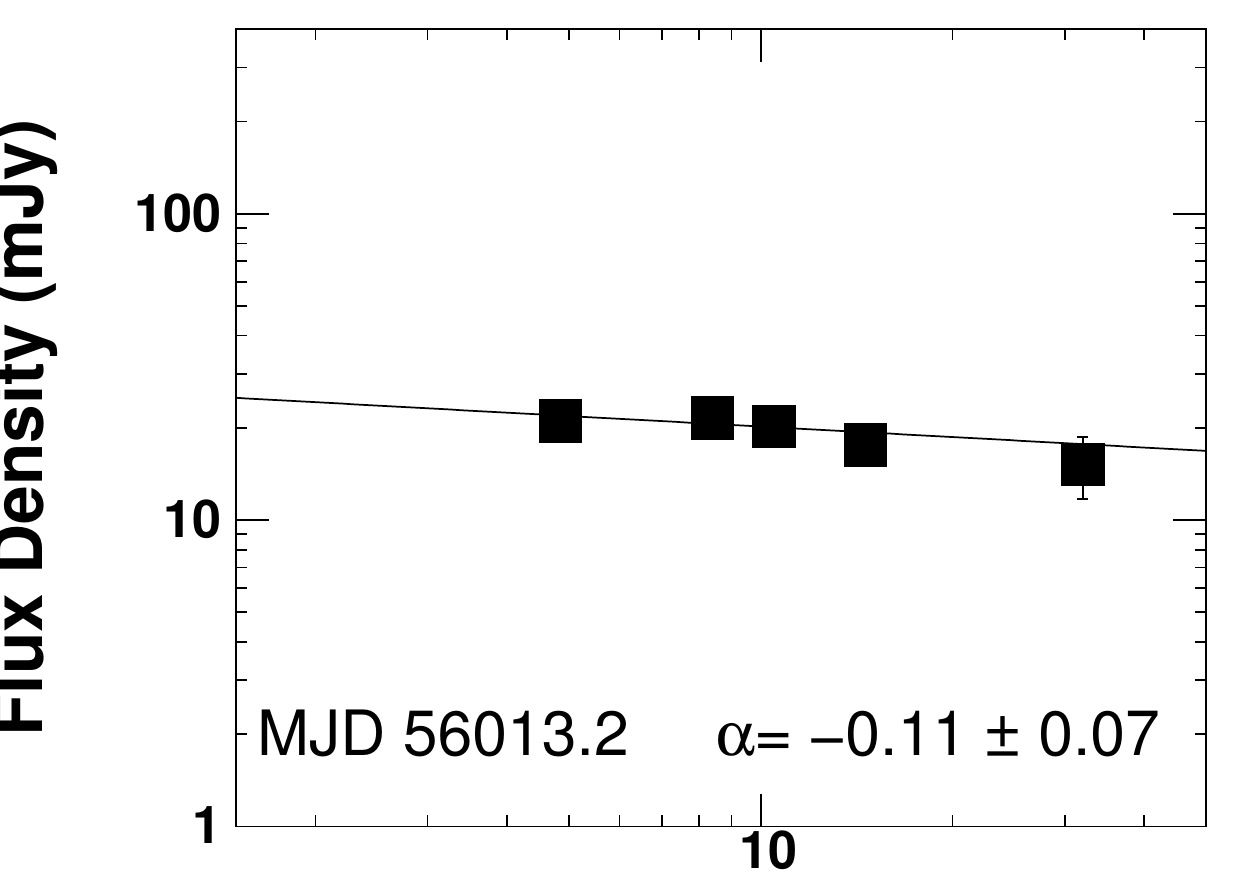}
\includegraphics[width=3.4cm,angle=0]{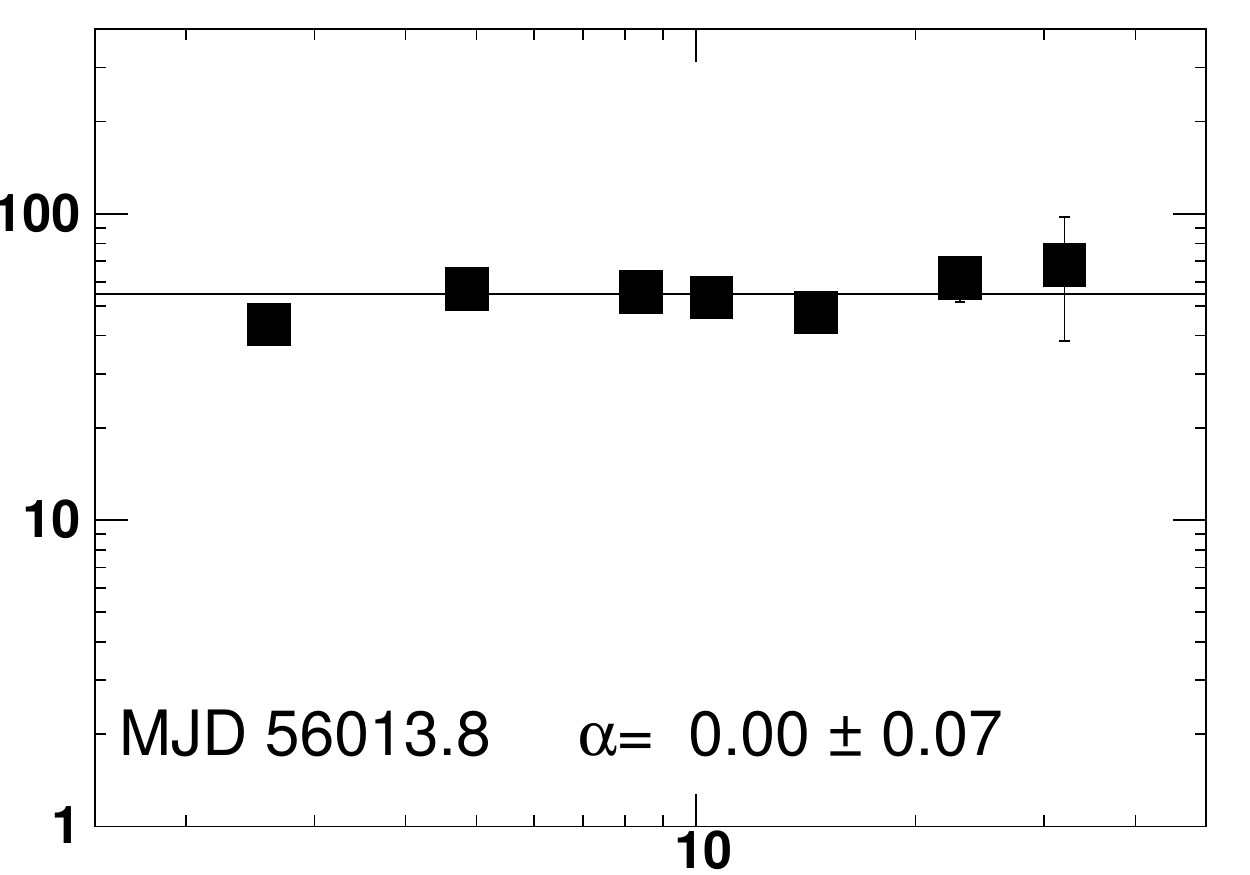}
\includegraphics[width=3.4cm,angle=0]{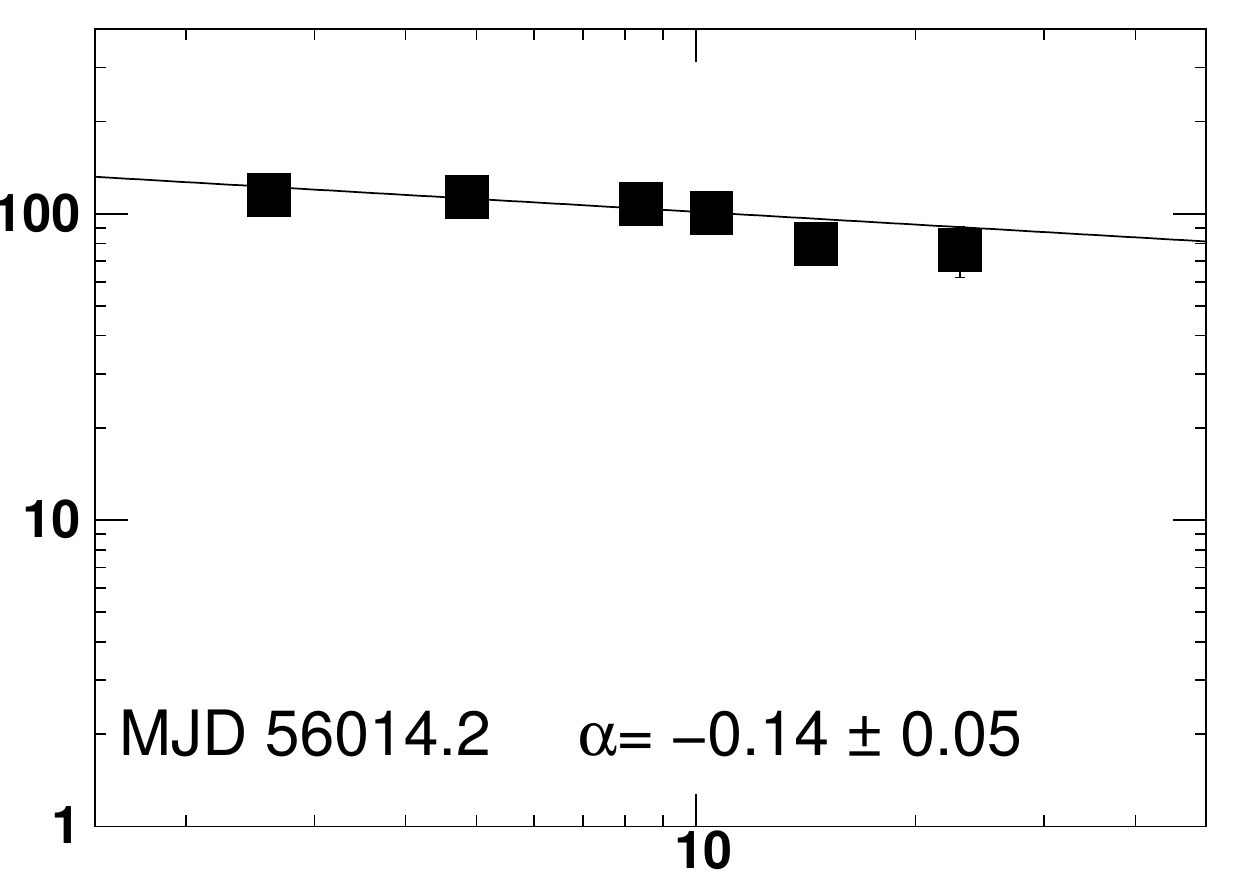}
\includegraphics[width=3.4cm,angle=0]{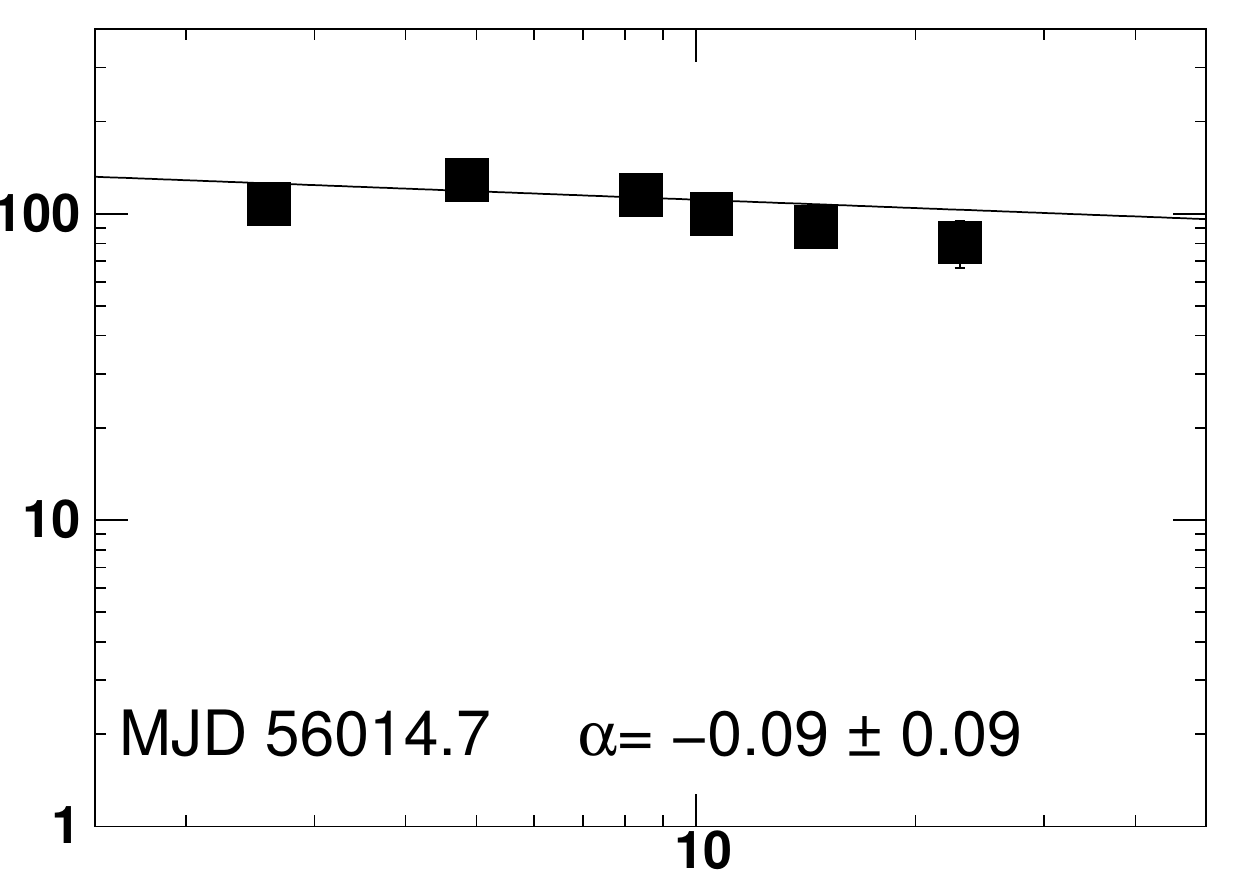}\\
\includegraphics[width=3.4cm,angle=0]{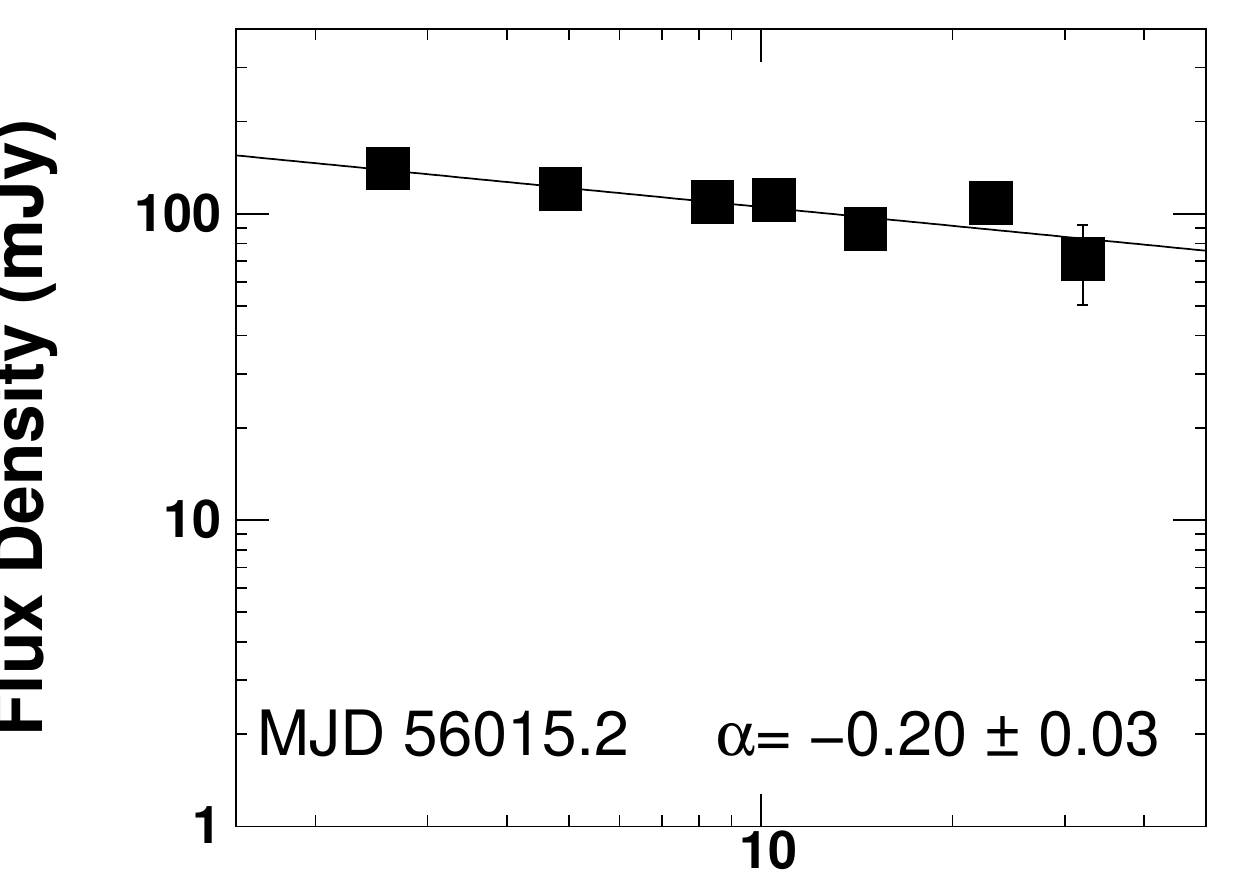}
\includegraphics[width=3.4cm,angle=0]{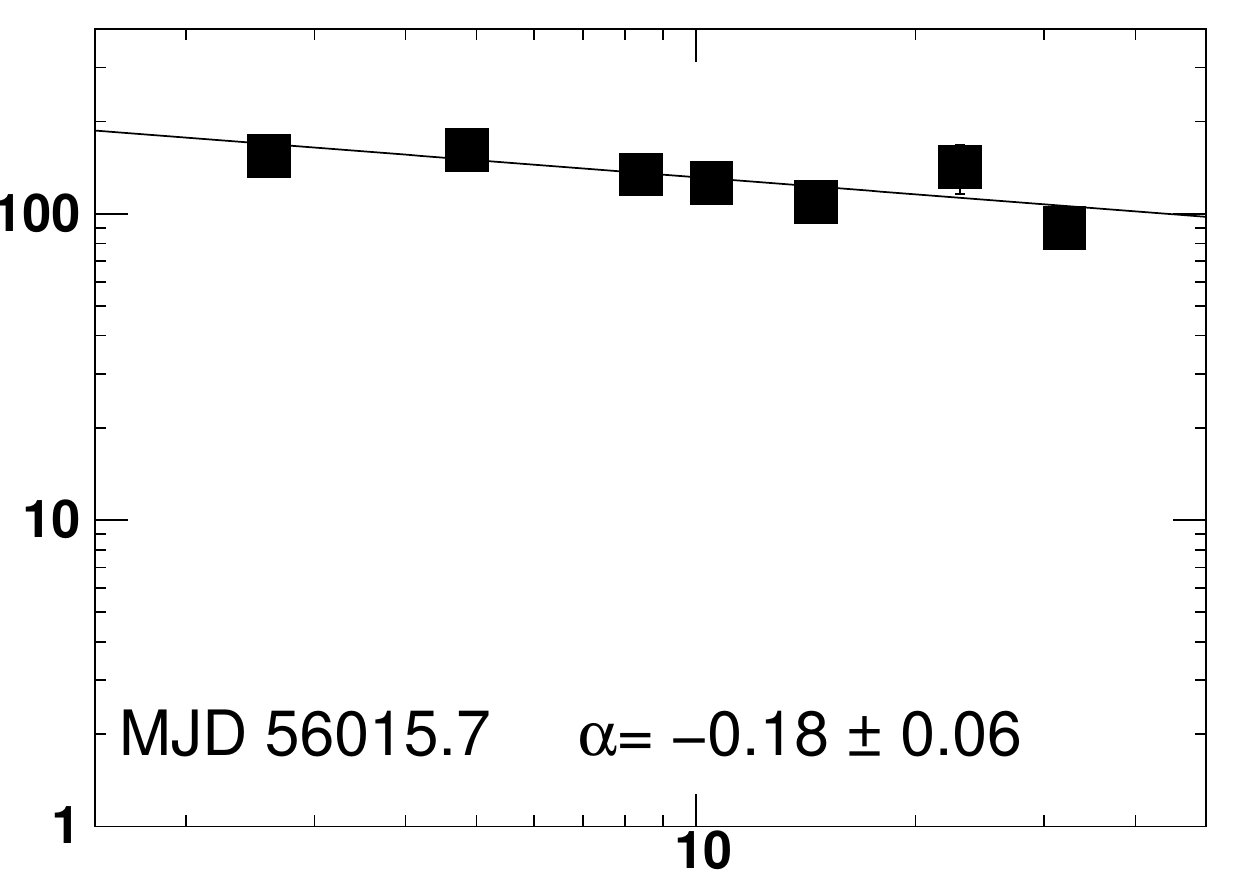}
\includegraphics[width=3.4cm,angle=0]{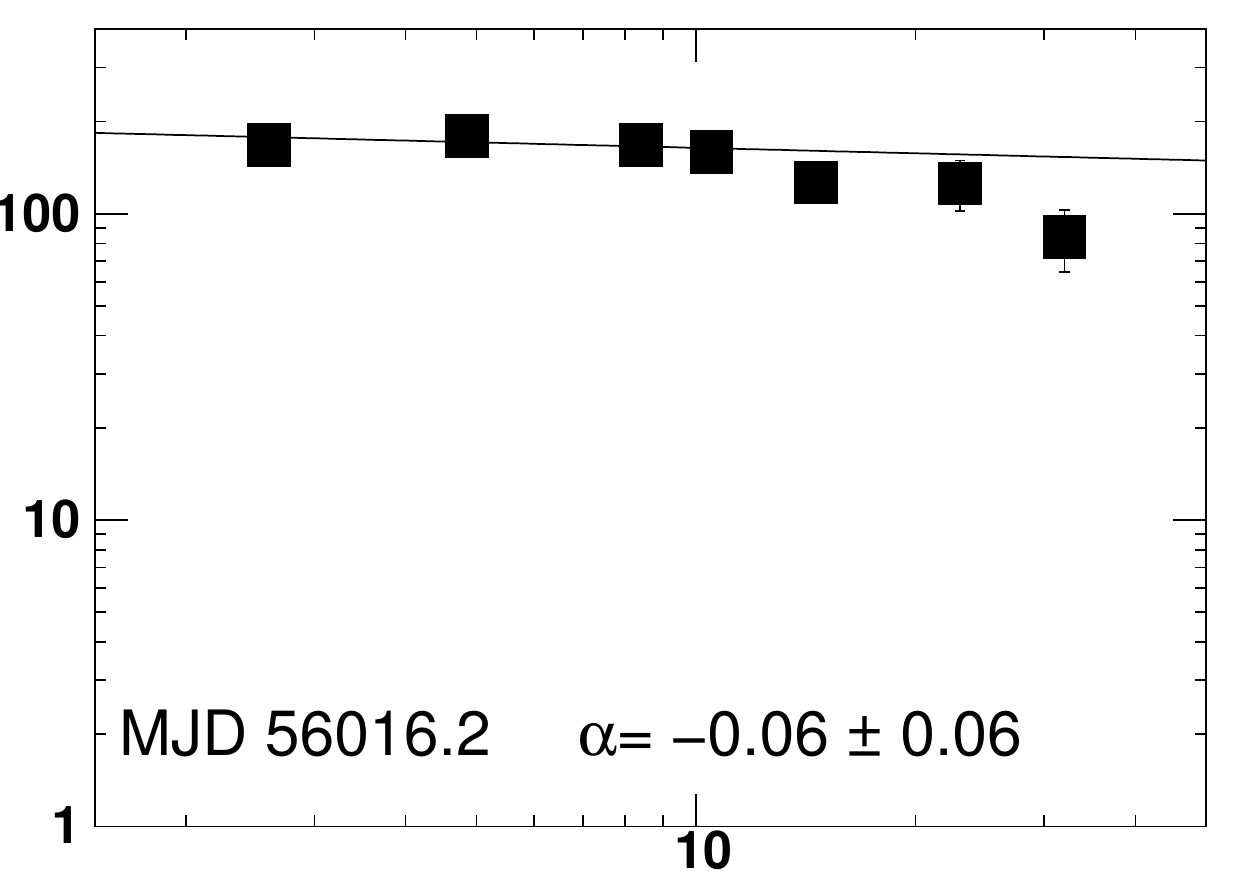}
\includegraphics[width=3.4cm,angle=0]{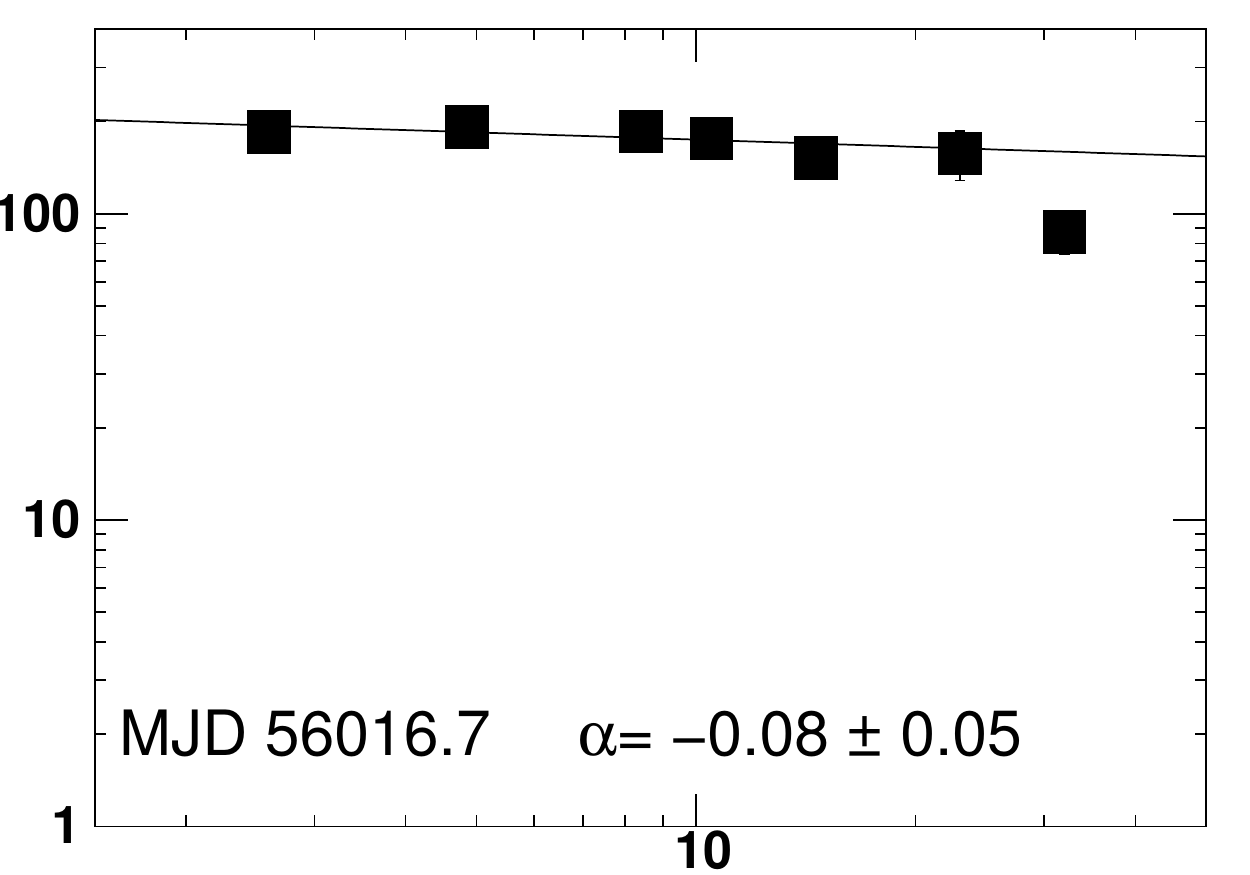}\\
\includegraphics[width=3.4cm,angle=0]{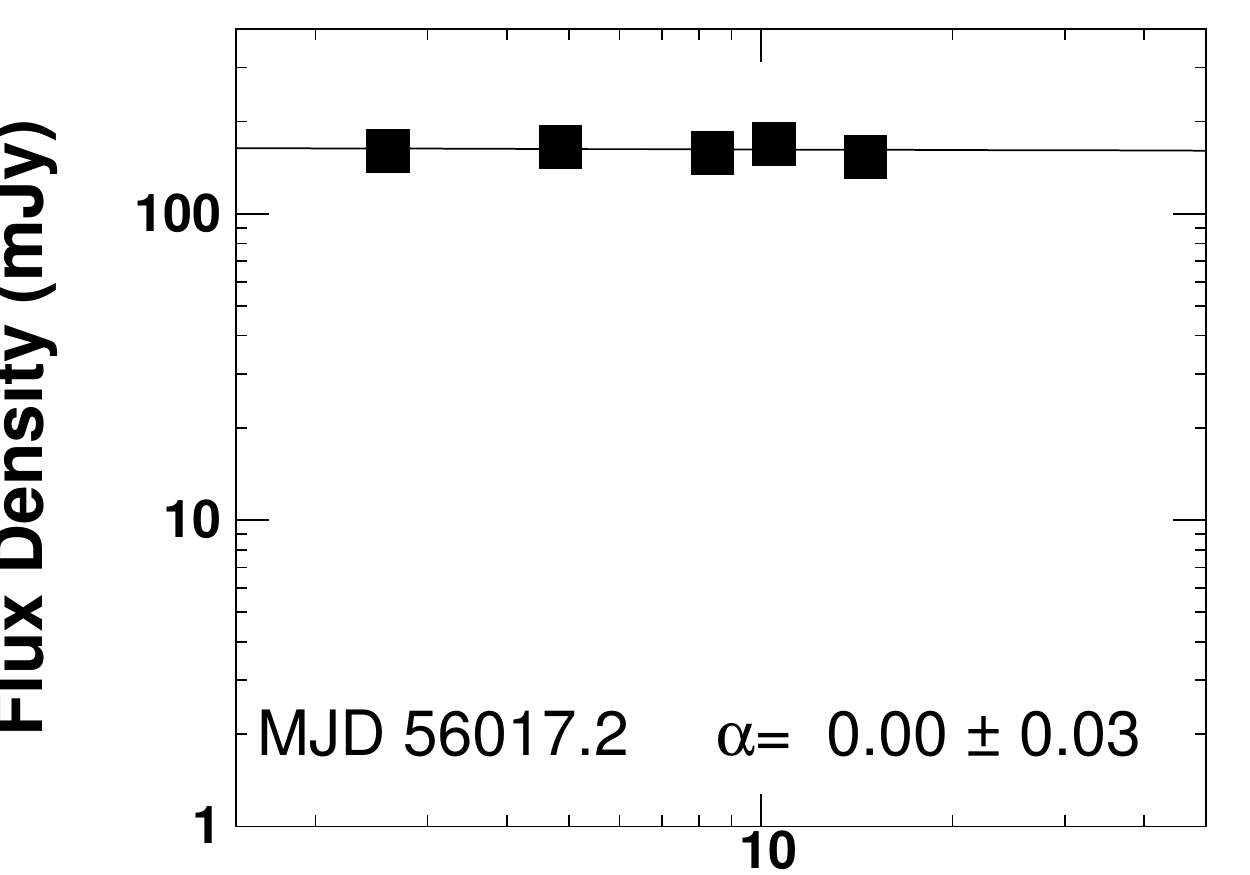}
\includegraphics[width=3.4cm,angle=0]{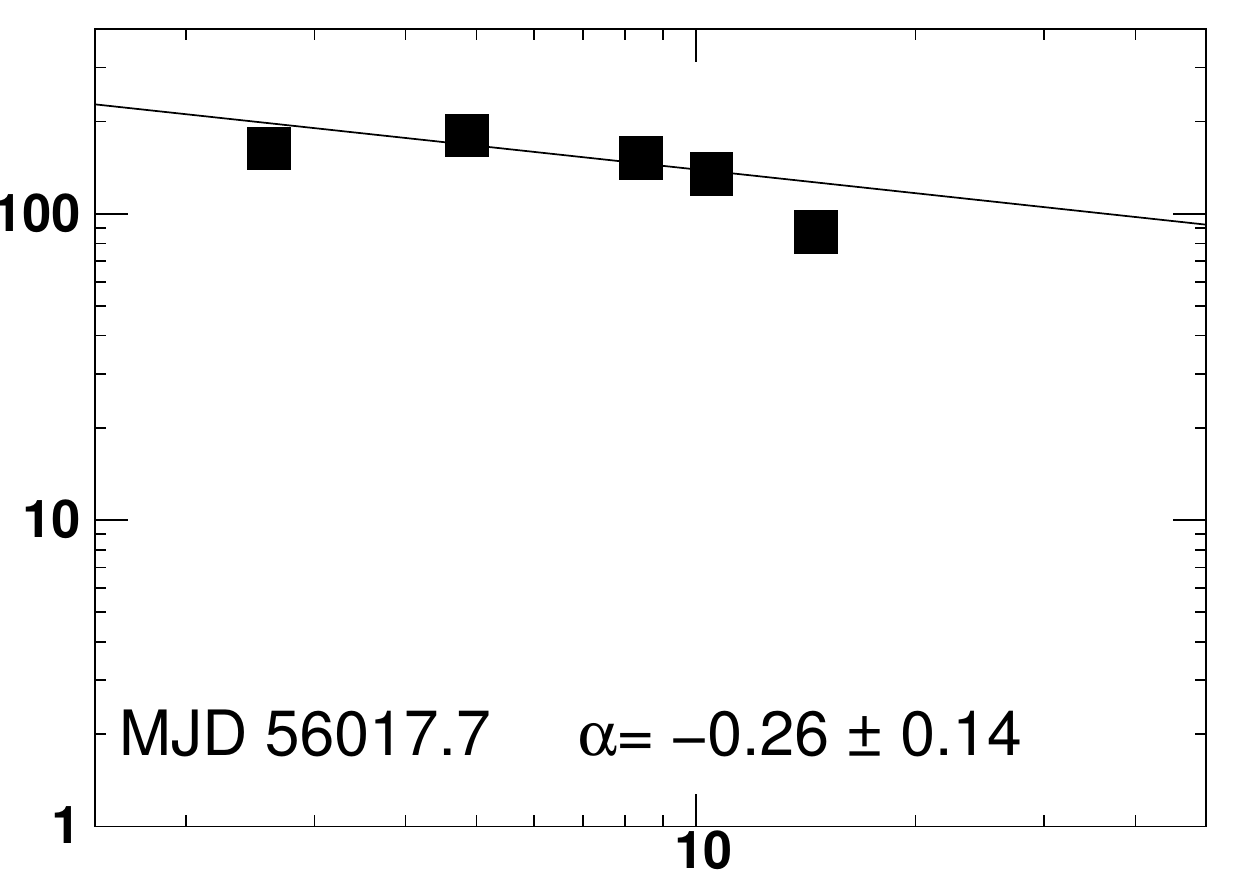}
\includegraphics[width=3.4cm,angle=0]{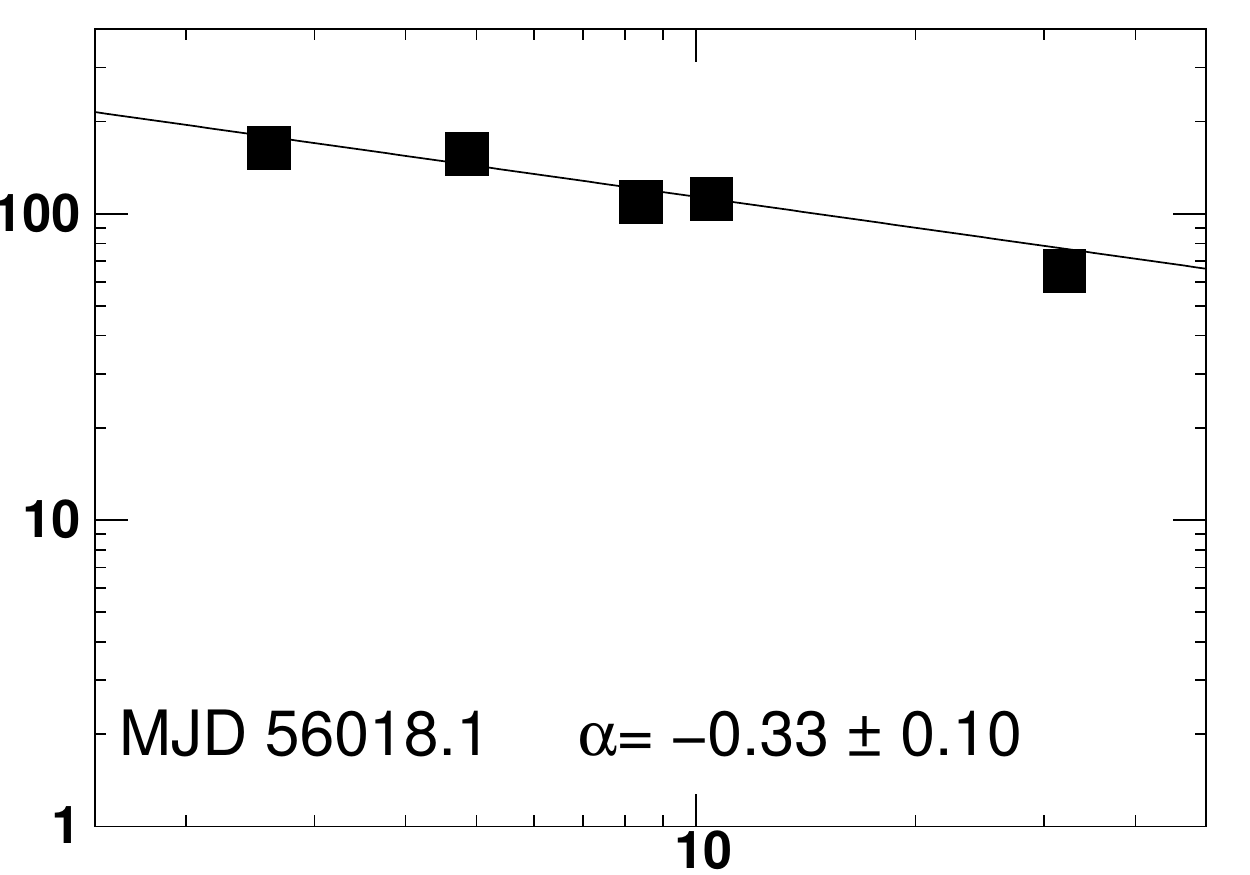}
\includegraphics[width=3.4cm,angle=0]{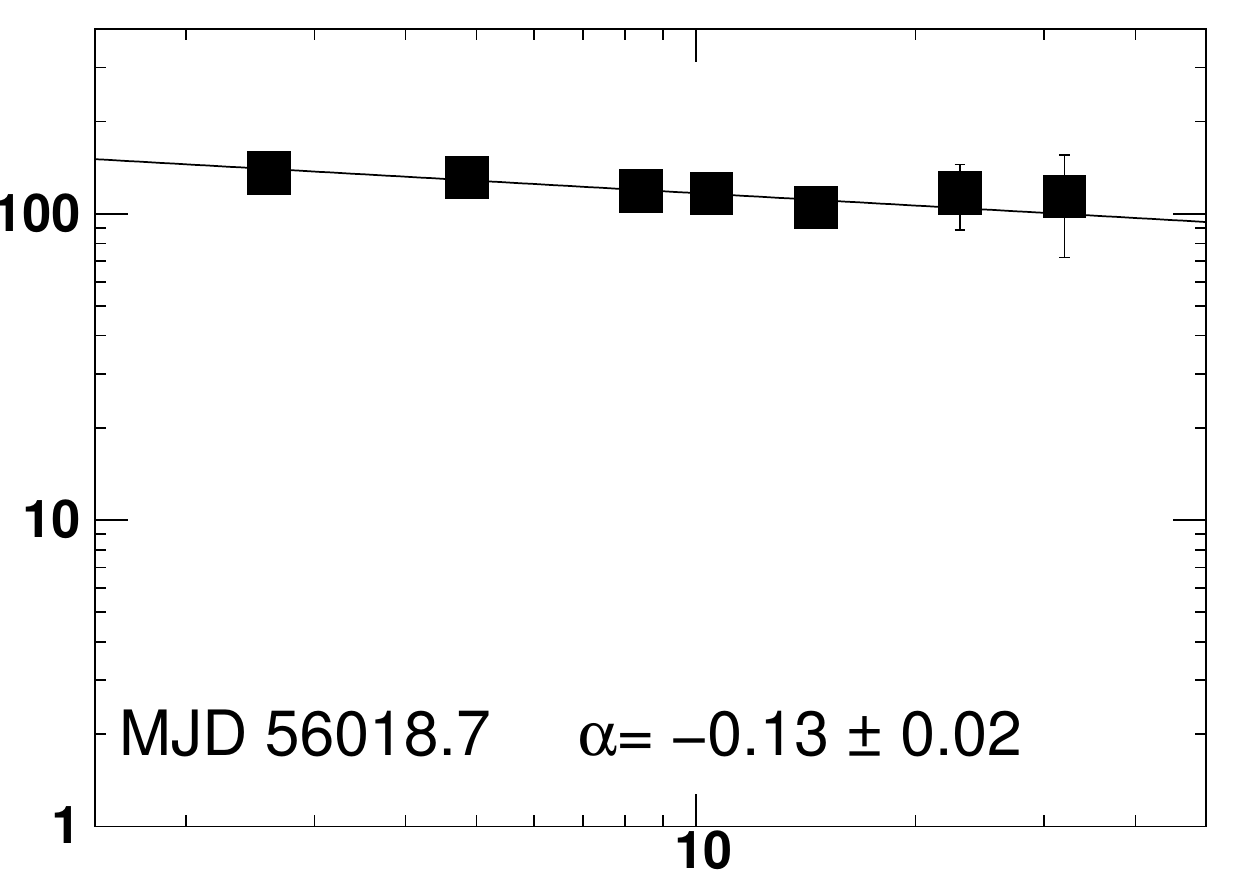}\\
\includegraphics[width=3.4cm,angle=0]{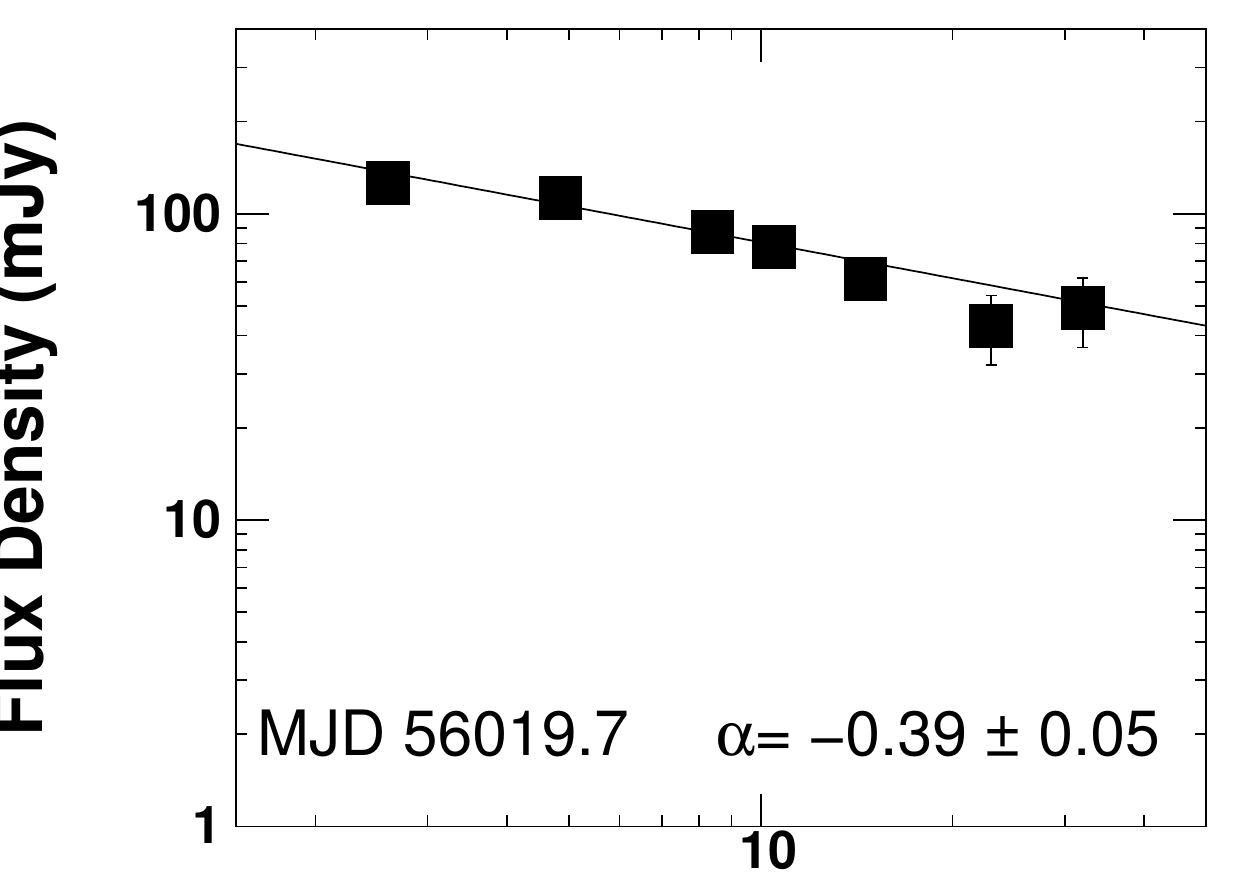}
\includegraphics[width=3.4cm,angle=0]{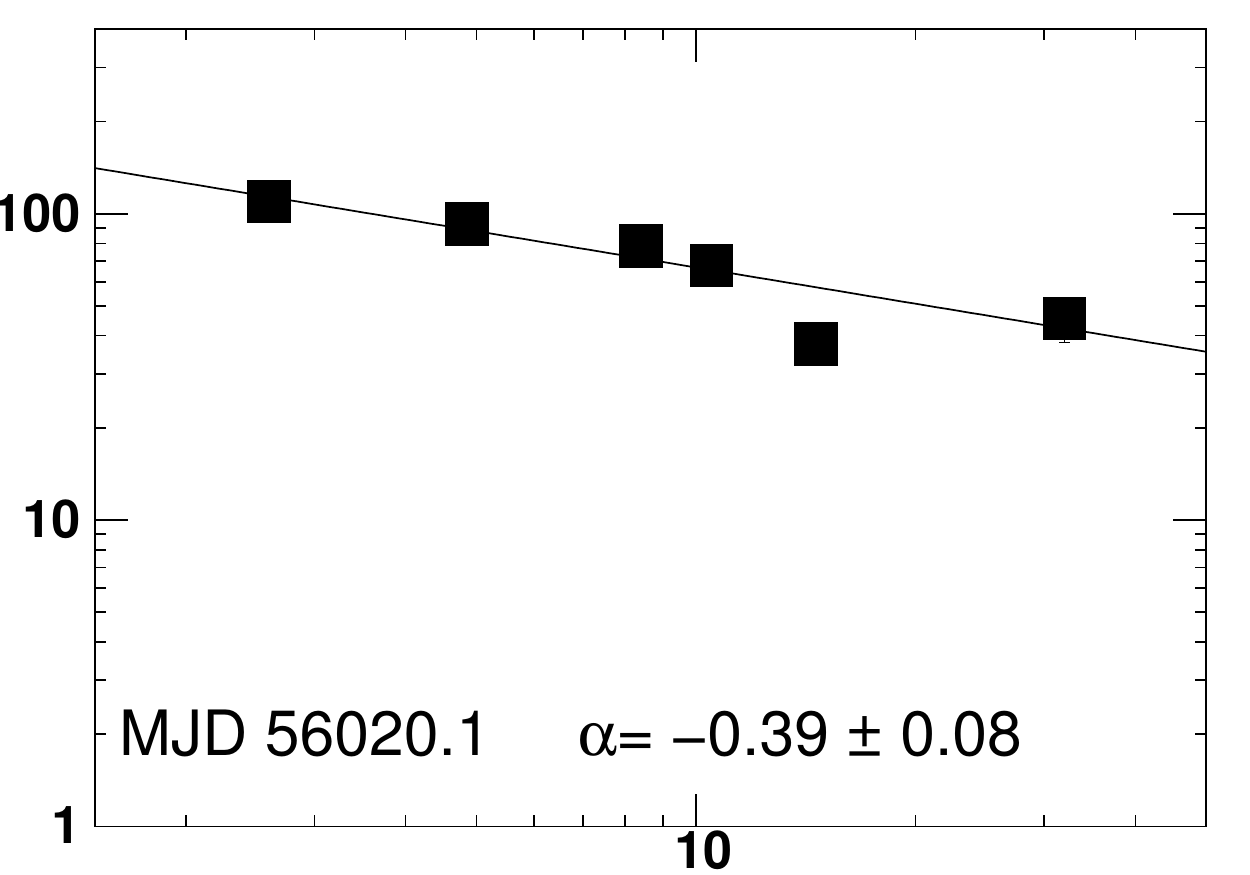}
\includegraphics[width=3.4cm,angle=0]{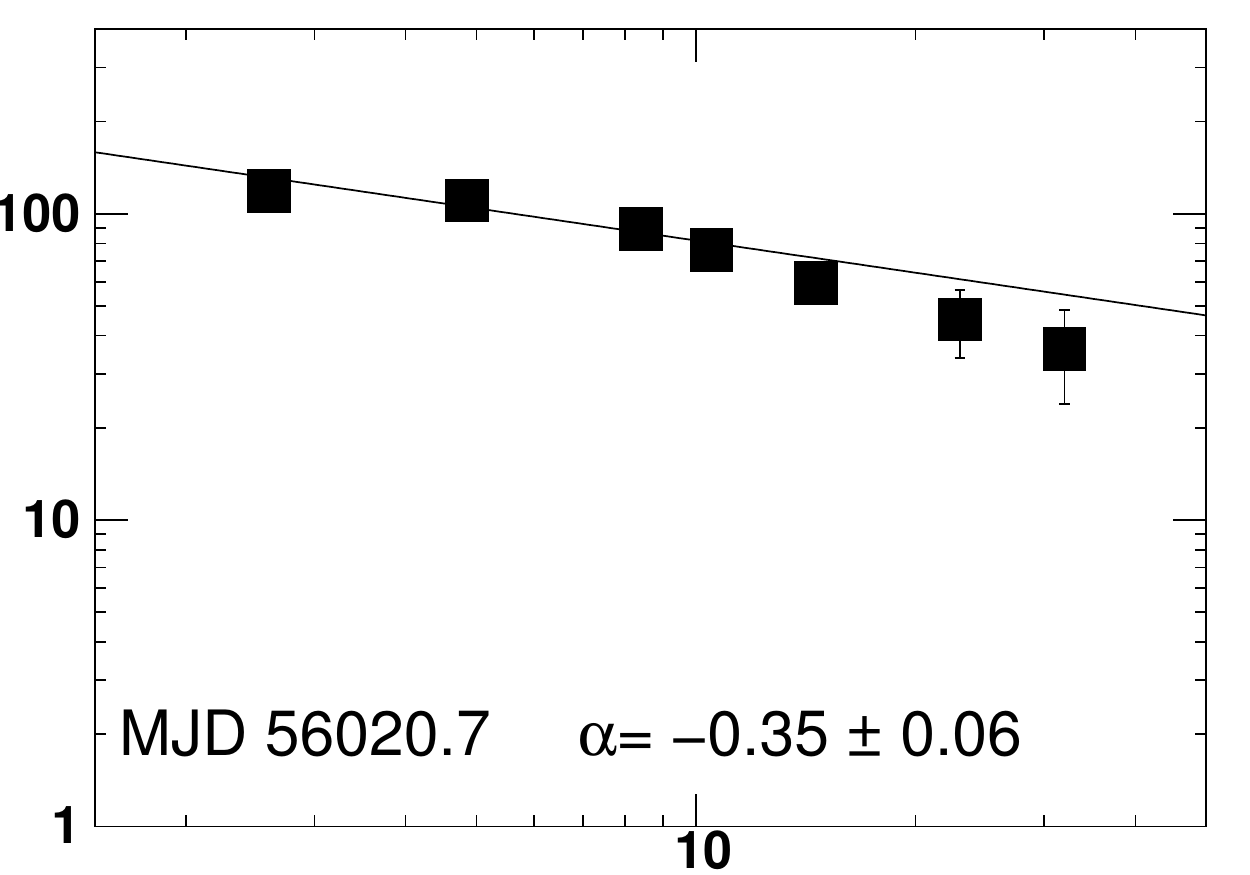}
\includegraphics[width=3.4cm,angle=0]{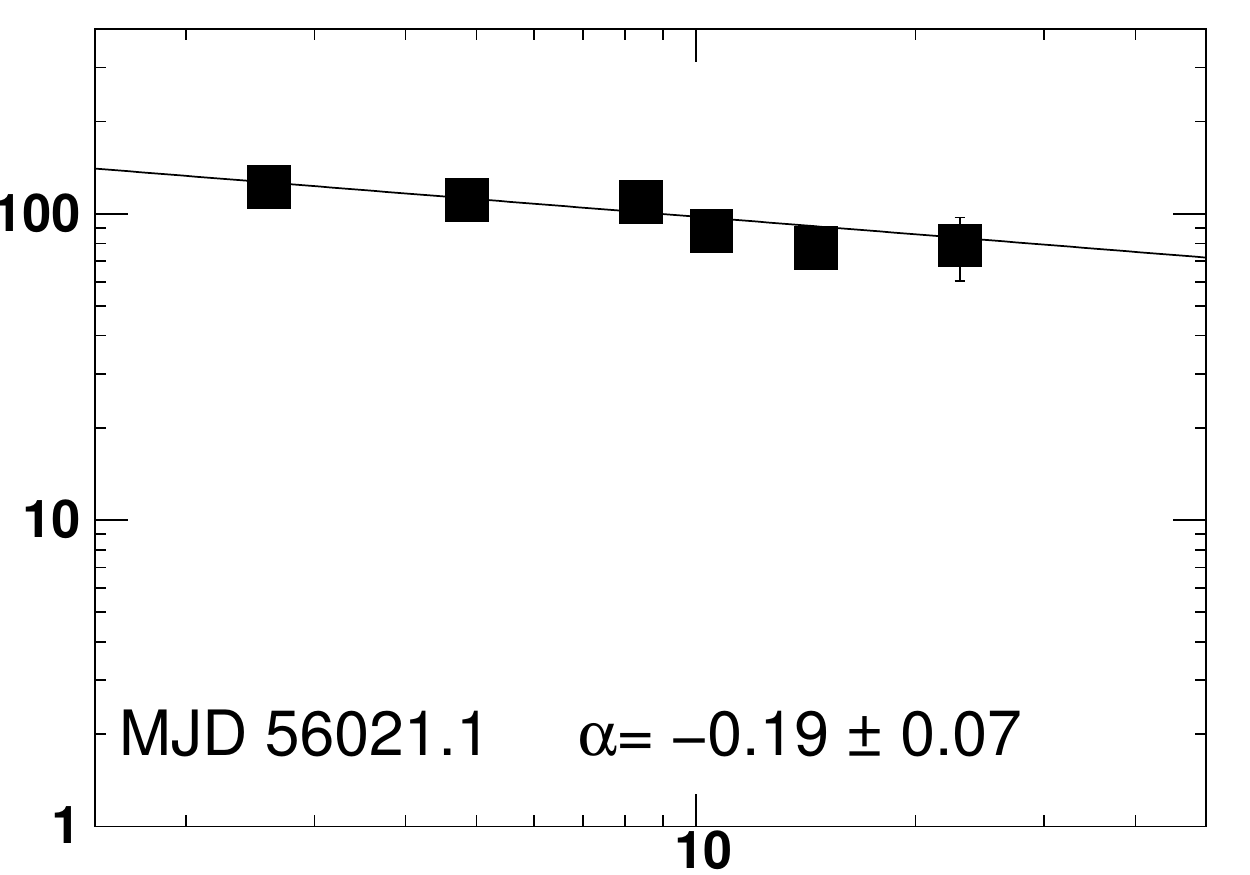}\\
\includegraphics[width=3.4cm,angle=0]{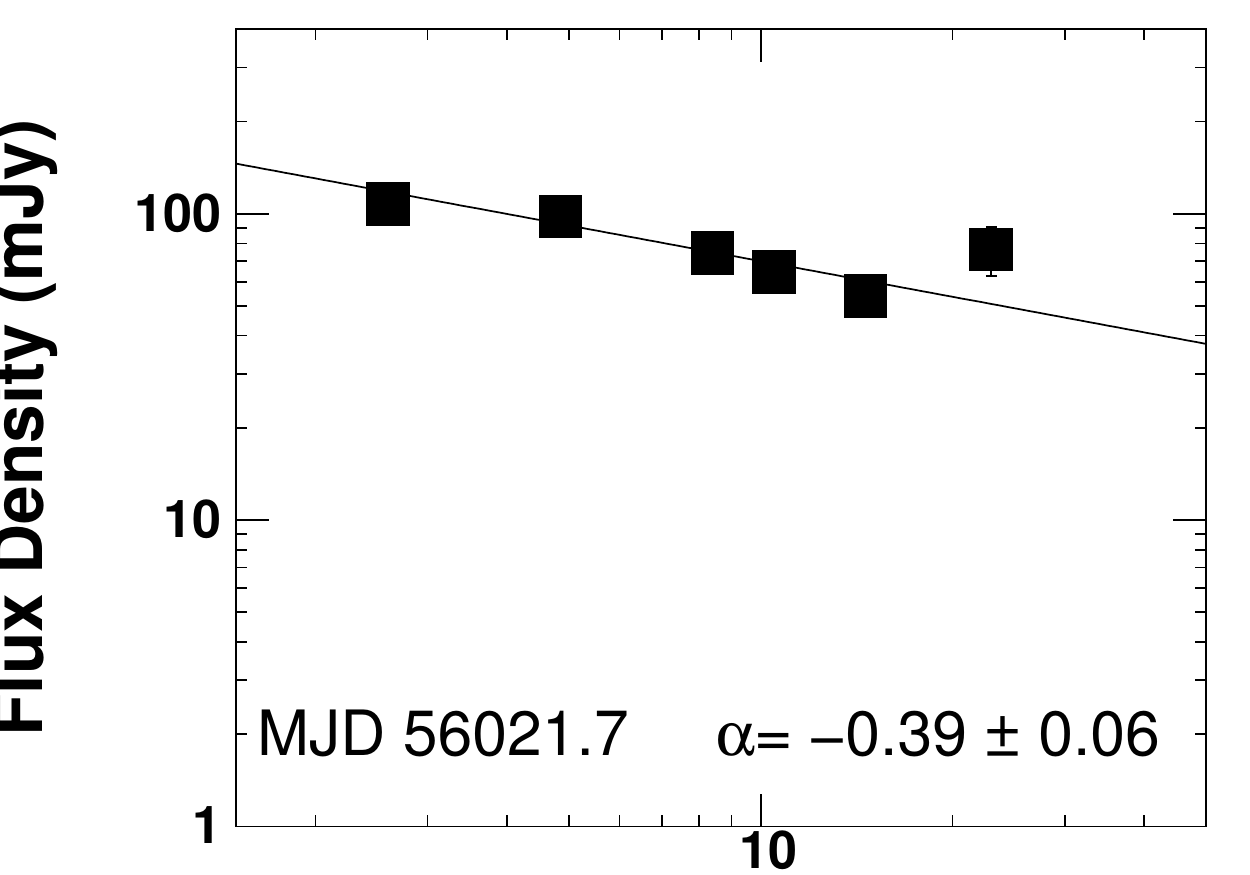}
\includegraphics[width=3.4cm,angle=0]{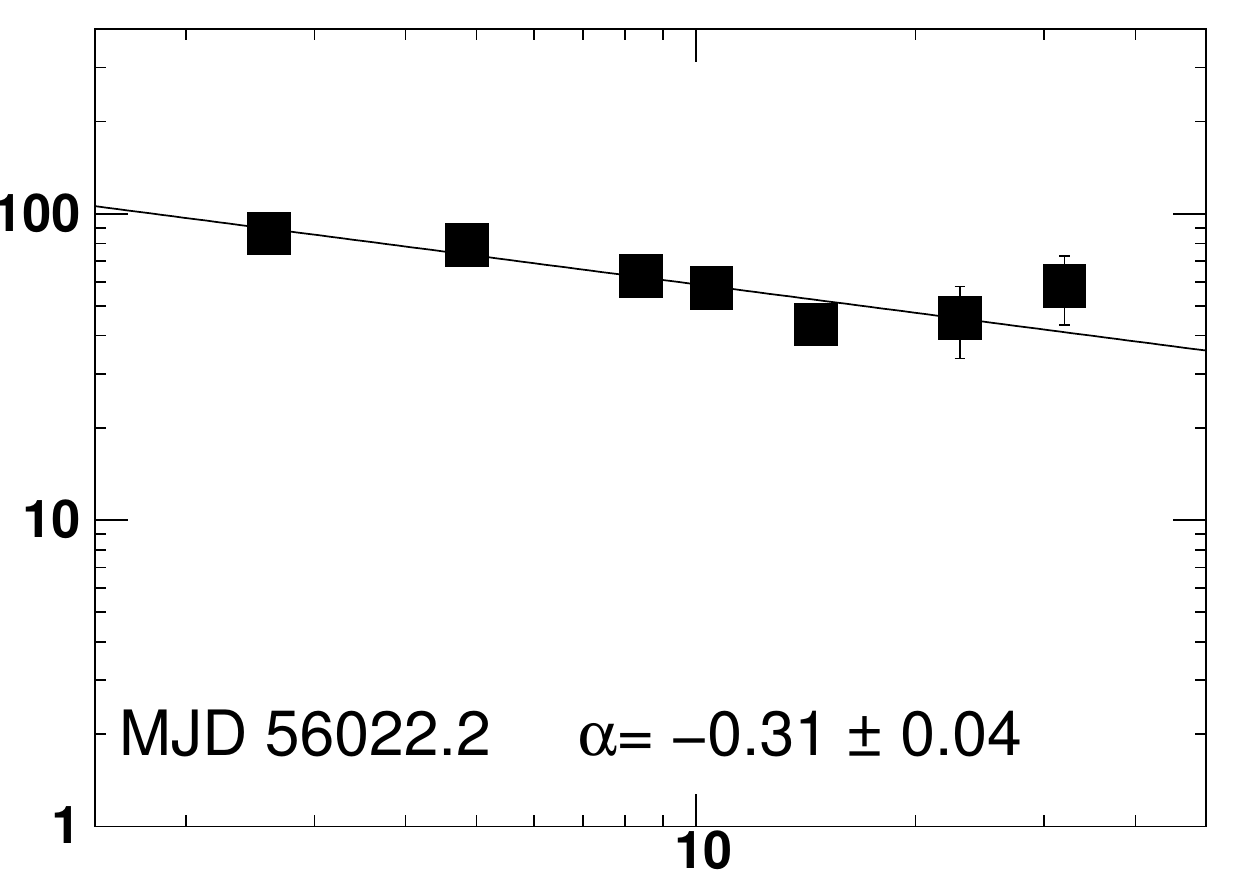}
\includegraphics[width=3.4cm,angle=0]{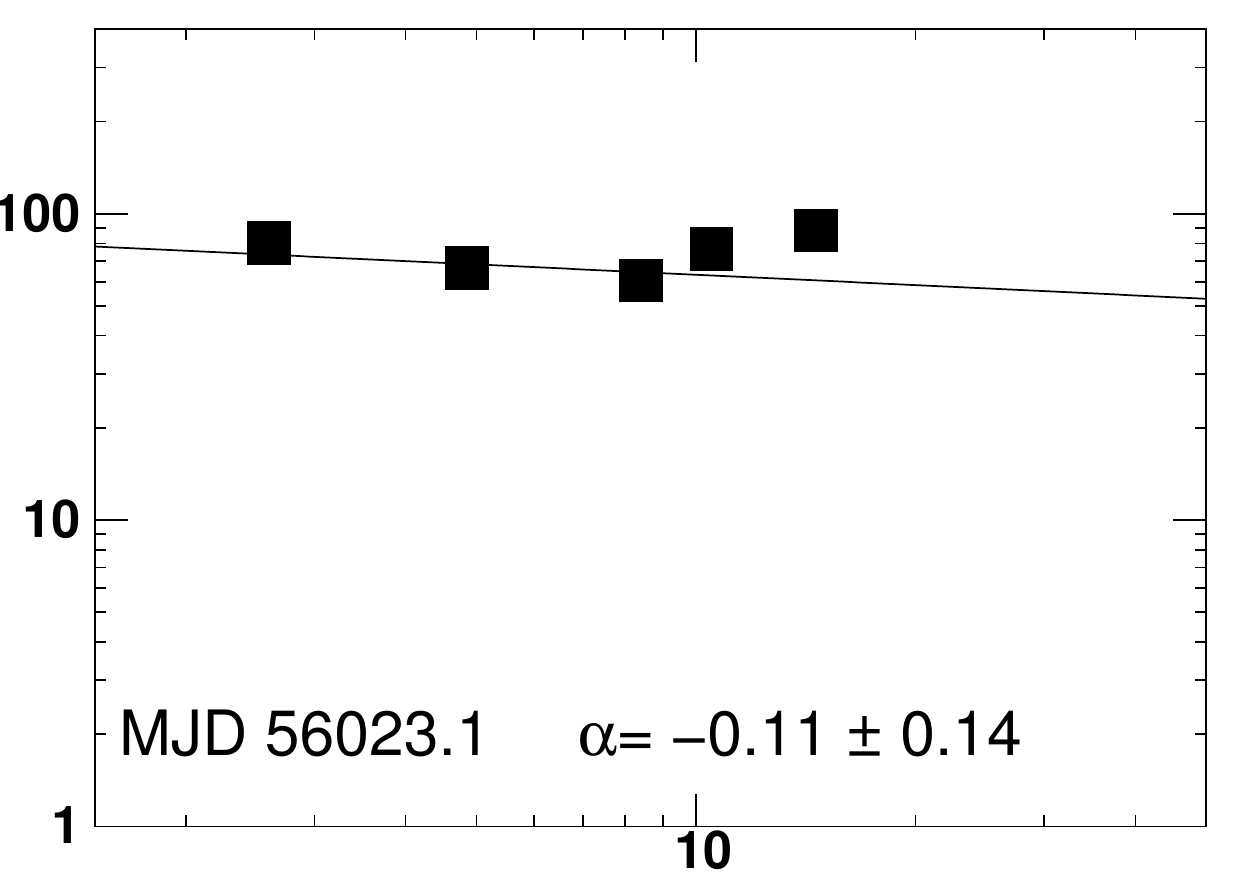}
\includegraphics[width=3.4cm,angle=0]{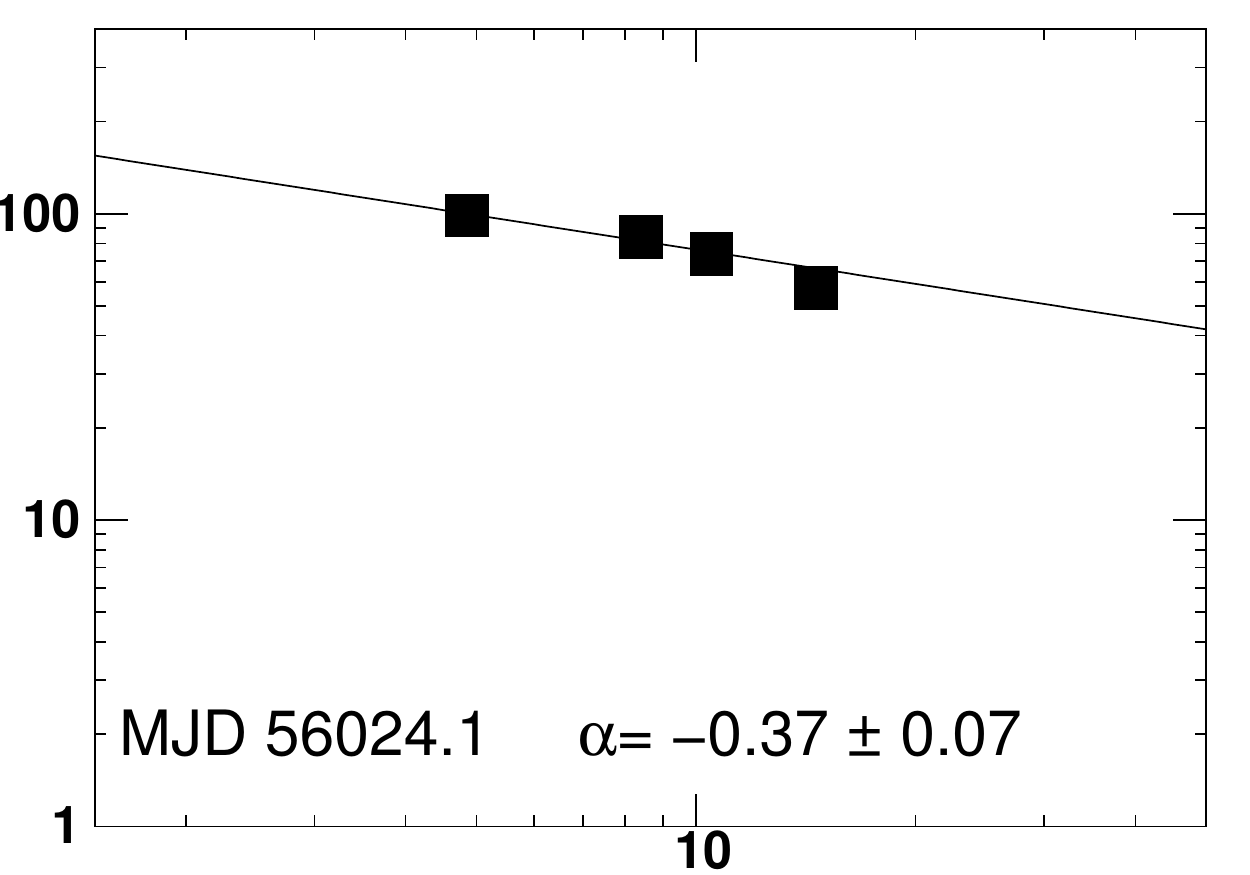}\\
\includegraphics[width=3.4cm,angle=0]{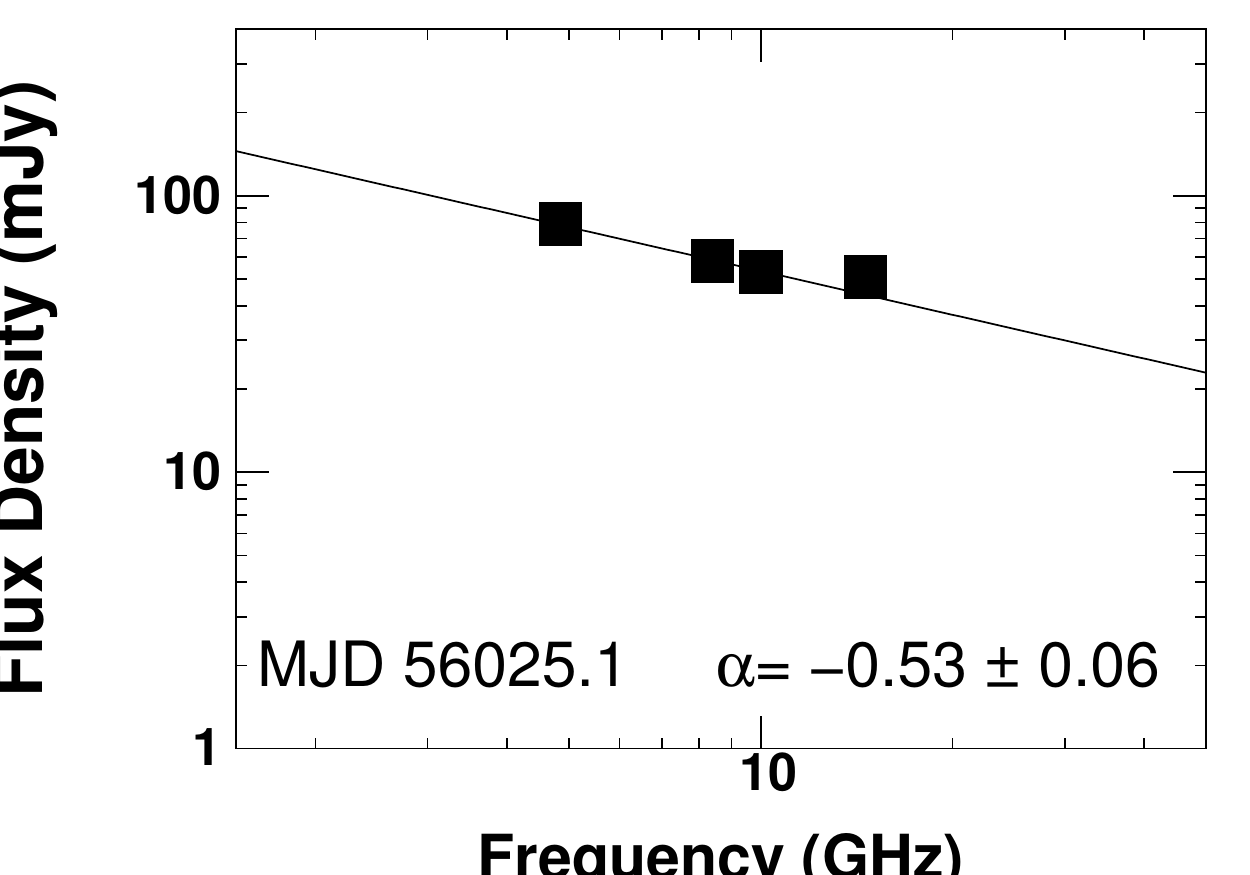}
\includegraphics[width=3.4cm,angle=0]{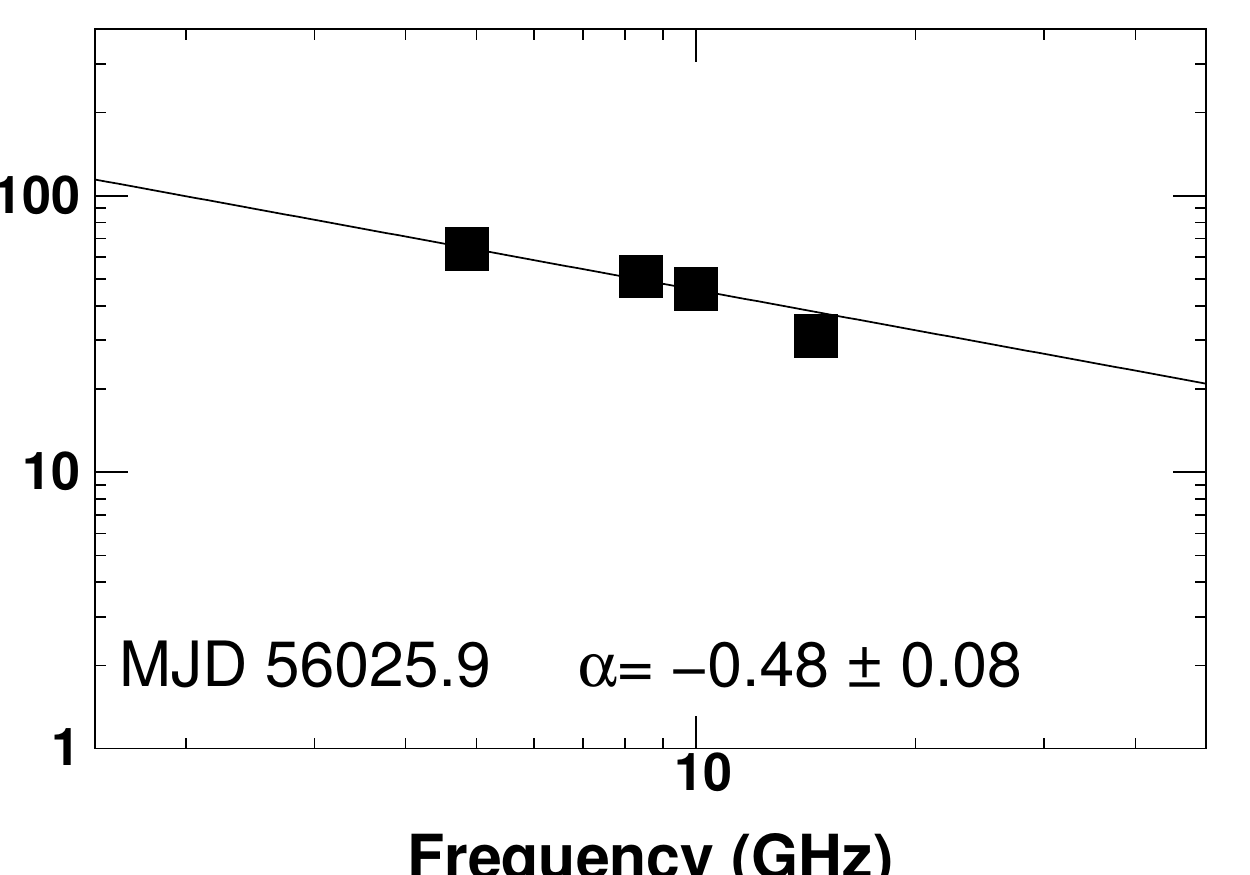}
\includegraphics[width=3.4cm,angle=0]{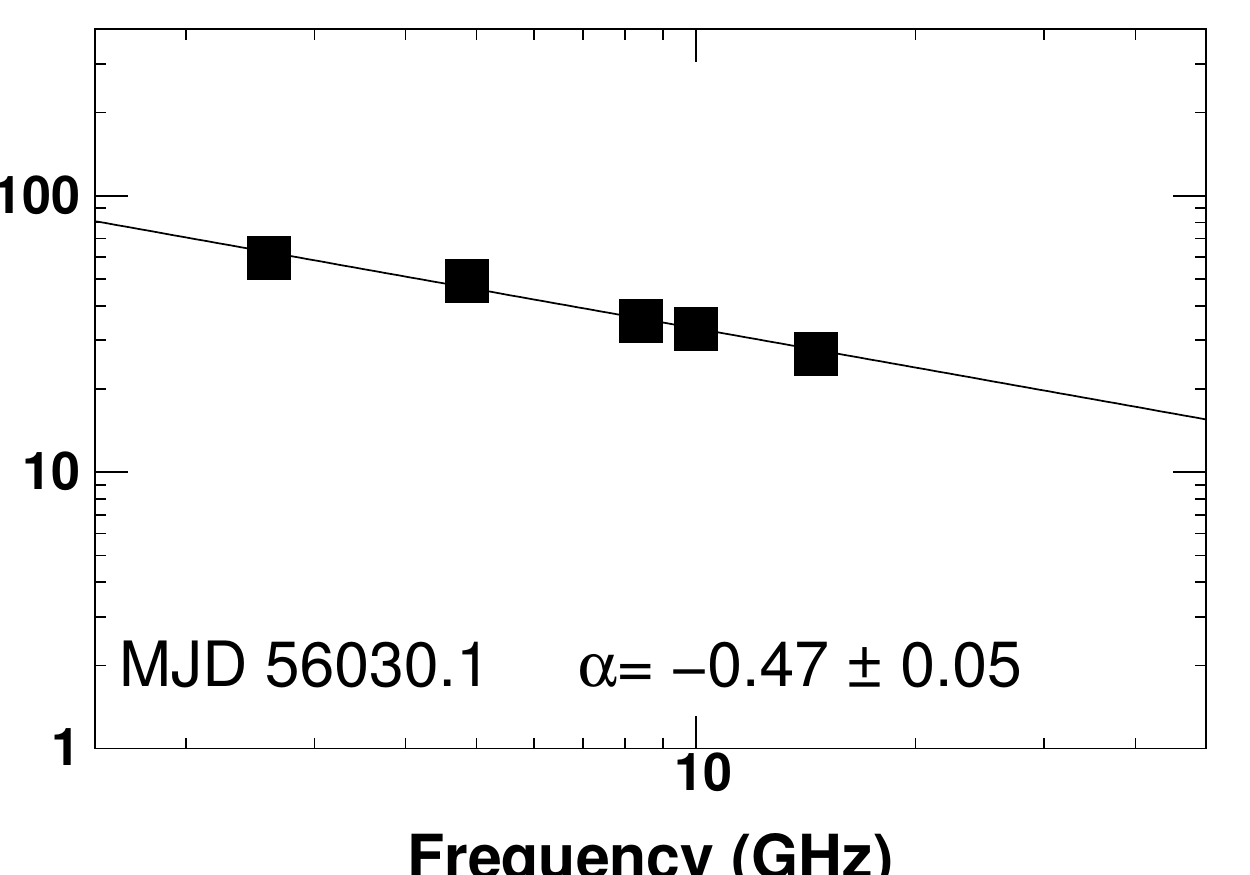}
\includegraphics[width=3.4cm,angle=0]{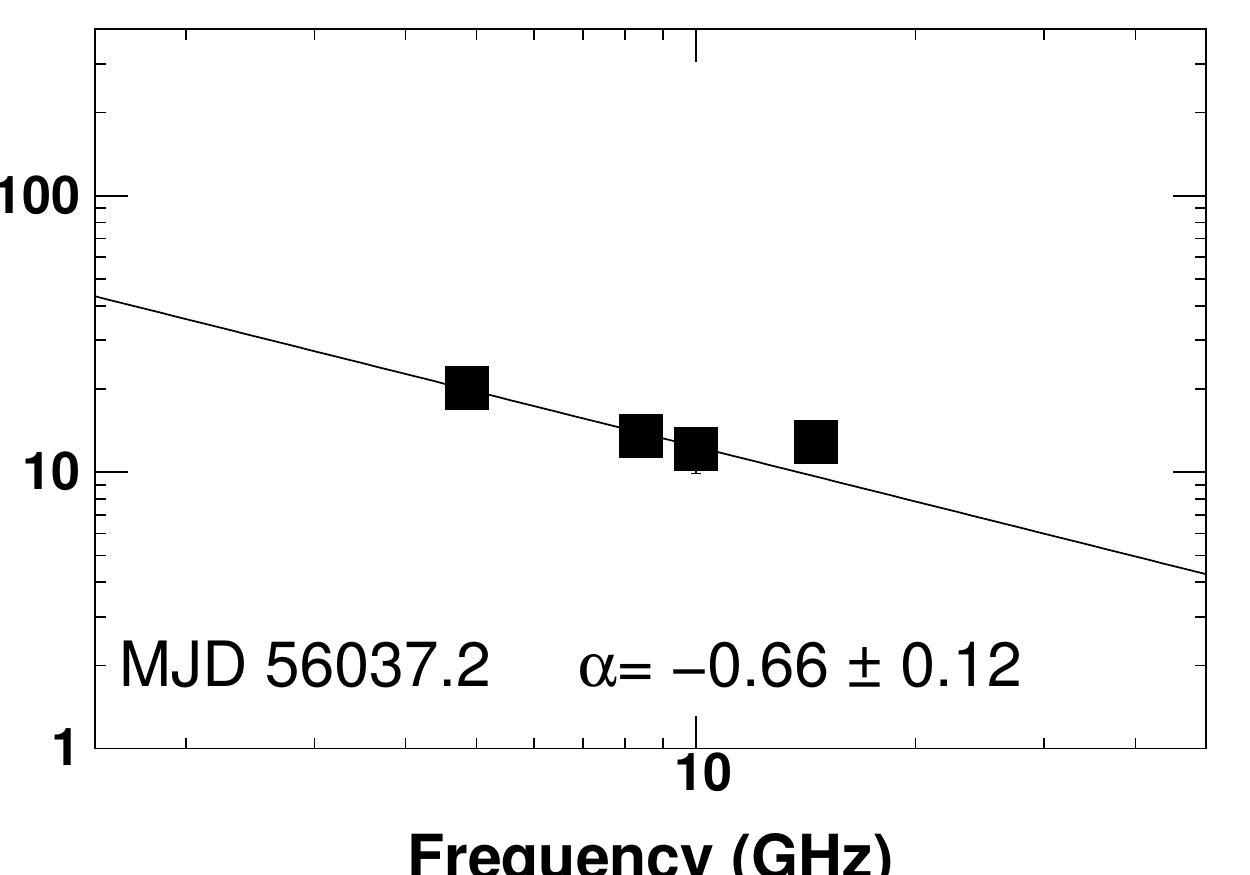}
\caption{Quasi-simultaneous broad-band radio continuum spectra
  obtained with the Effelsberg 100 m telescope.  The spectral index
  $\alpha$ resulting from a power-law fit is indicated in each panel
  together with the time of observation. The measurement errors are
  comparable to or smaller than the symbol size. We note that up to MJD
  56017.2 the spectral index is flat with an average of $-0.10\pm0.02$
  (see text).}.
\label{spectra}
\end{figure*}

\section{Observations and data reduction}
The multi-frequency flux density measurements of the periodical
\citep{jaronmassi13} radio outburst of \lsi were performed
approximately every 12\,hours over 14 consecutive days between 2012
March 27, orbital phase $\Phi=0.31$, and 2012 April 8, $\Phi=0.78$
\citep{zimmermann13}.  The orbital phase $\Phi$ is equal to the
fractional part of $(t-43366.275)/P_{\rm orbital}$ \citep[$P_{\rm
  orbital}=26.4960\pm 0.0028$ d,][]{gregory02}. Two additional
measurements were obtained on 2012 April 12 ($\Phi=0.94$) and 19
($\Phi=0.2$) bringing the total observing period to 24\,days.  At each
observing session lasting for about 2 hours, \lsi was observed
together with primary calibrators consecutively at seven frequencies,
namely 2.64, 4.85, 8.35, 10.45, 14.60, 23.02, and 32.00\,GHz (110, 60,
36, 28, 20, 13, and 9\,mm wavelengths) using the secondary focus
receivers of the Effelsberg 100 m telescope. Flux density measurements
were performed using the cross-scan technique, i.e.  progressively
slewing over the source position in azimuthal and elevation direction
with the number of sub-scans matching the source brightness at a given
frequency.  The data reduction from raw telescope data to calibrated
flux densities/spectra was done in the standard manner as described in
\citet{fuhrmann08}. The overall, final measurement uncertainties are
of the order of $\le$\,1\% and $\le$\,5\% at lower and higher
frequencies, respectively.

\section{Results}

The total flux density evolution at all frequencies is shown in
Fig. \ref{LCs} (top). At MJD 56013.2 ($\Phi$=0.29), the flux density
rises steeply from around 20 mJy to 100 mJy within 24 hours (until MJD
56014.2) and remains approximately at that level. After another 24
hours (MJD 56015.7), it rises yet again to the highest flux density
level (around 200 mJy at the lowest frequencies) at MJD 56016.7,
i.e. the outburst reaches its peak within about 3.5\,days after
onset. We further note that the peak of the outburst at frequencies
$\le$\,10.45\,GHz is reached at the same epoch, i.e.  no significant
delay is obvious within our time sampling of $\sim$\,12\,hours. At
higher frequencies the situation is less clear given the lower data
sampling and larger measurement uncertainties.  Subsequently, the slow
decay of the outburst takes place (MJD\,$\ge$\,56017.2) though
intersected by several superimposed events. These sub-flares occur
with rise/decay times of $\sim$\,12 to 24\,hours and with amplitudes
of a few 10\,mJy.  We note that during the last sub-flare (around MJD
56024), the peak at higher frequencies (here: 10.45 and 14.60\,GHz)
appears to occur 24\,hours earlier.  In Fig. \ref{spectra}, the
broad-band radio spectra of all epochs are shown. In order to quantify
the spectral behaviour  of \lsp, broad-band spectral indices $\alpha$
($S\propto\nu^{\alpha}$) were obtained by performing a power-law fit
to all available multi-frequency flux densities of a given epoch and
the resulting fits are superimposed in Fig. \ref{spectra}. The
spectral index evolution with time is shown in Fig. \ref{LCs} (bottom:
$\alpha$ vs. MJD) and is characterized by an overall long-term
decreasing trend with (i) the flattest spectral indices occurring
during the early raise/peak of the outburst and (ii) increasingly
steeper values during the outburst decay with the steepest values
reached towards the end ($\alpha\sim$\,$-$0.5 to $-$0.7).  A linear
fit to the spectral indices reveals that on average a spectral
steepening of $\Delta\alpha$\,=\,0.03 per day takes place over the
total observing period of 24\,days.  Examining the spectral index
behaviour in more detail we basically identify three stages (vertical
dotted lines in Fig. \ref{LCs}, bottom).  During the rise until
shortly after the main peak, i.e. from MJD 56013.2 to MJD 56017.2, the
broad-band spectrum remains almost flat (average spectral index
$\alpha=-$0.10$\pm$0.02) and an even inverted spectrum is observed at
high frequencies during some epochs (i.e. $\alpha_{\rm
  14~GHz/23~GHz}\sim$0.6, 0.4, and 0.6 at MJD 56013.8, MJD 56015.2, and
MJD56015.7, respectively).  Subsequently, a stage covering the
decay phase follows with the superimposed sub-flares lasting until MJD
56023.1 and  with spectral indices dropping to values as low as $-$0.39
(mean value: $-$0.29$\pm$0.03). Here, however, $\alpha$ appears to
oscillate between $-$0.4 and $-$0.1 in a quasi-regular fashion,
whereas the local peaks in spectral index roughly coincide with the
peaks of the sub-flares seen in the total intensity light curves
(Fig. \ref{LCs}, top). We further note that also during the decay stage
several epochs show an inverted spectrum at high frequencies. During
the last stage of the decay (MJD 56025 to MJD 56037), the spectral
index finally drops to the steepest observed values ranging between
$-$0.5 and $-$0.7, i.e. clearly optically thin, with an average of
$-0.5 \pm 0.1$ during this period.

\section{Discussion and conclusions}

\begin{figure}[t]
\centering
\includegraphics[width=4.45cm,  angle=0]{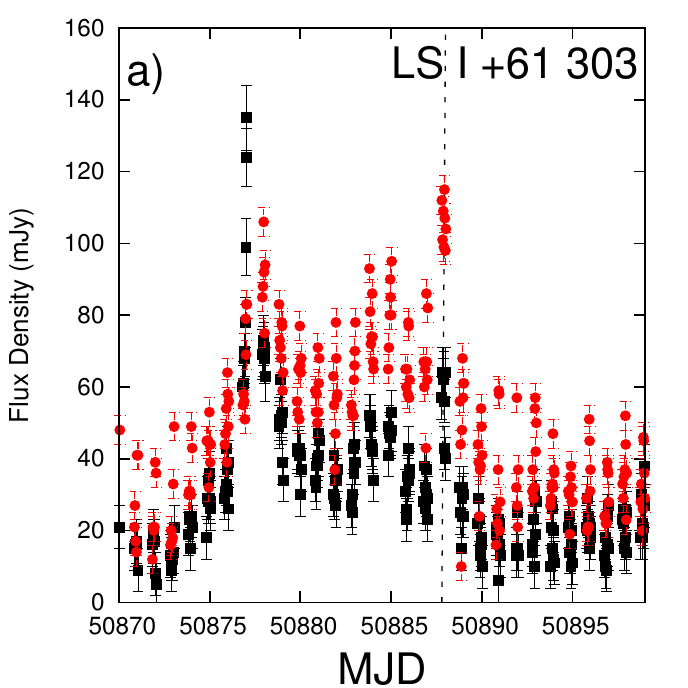}
\includegraphics[width=4.45cm,  angle=0]{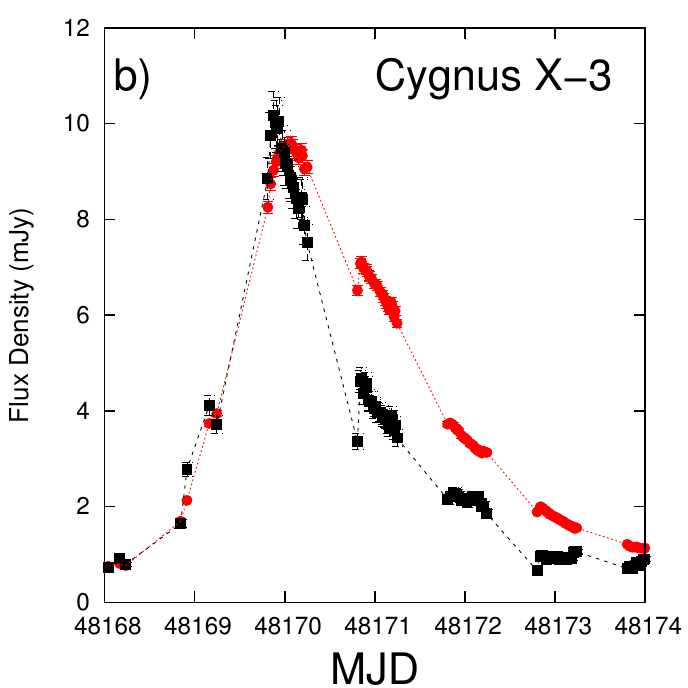}\\
\includegraphics[width=4.45cm,  angle=0]{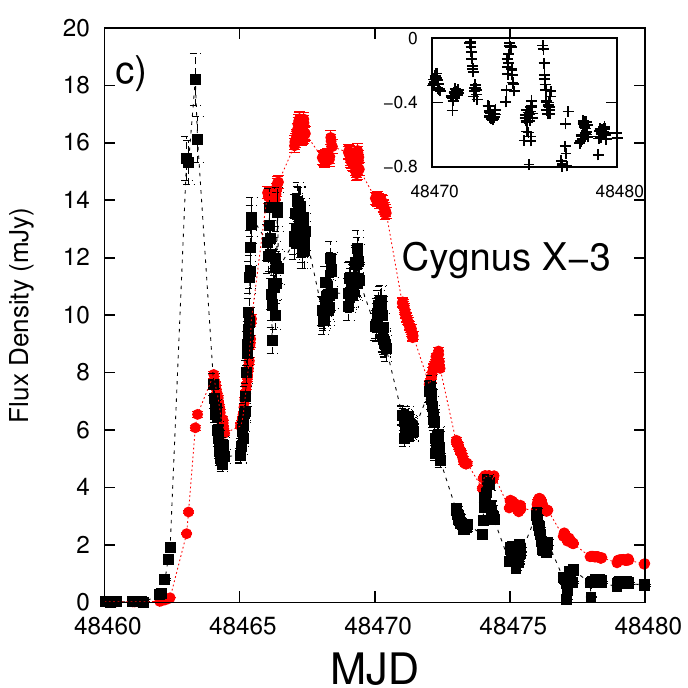}
\includegraphics[width=4.45cm,  angle=0]{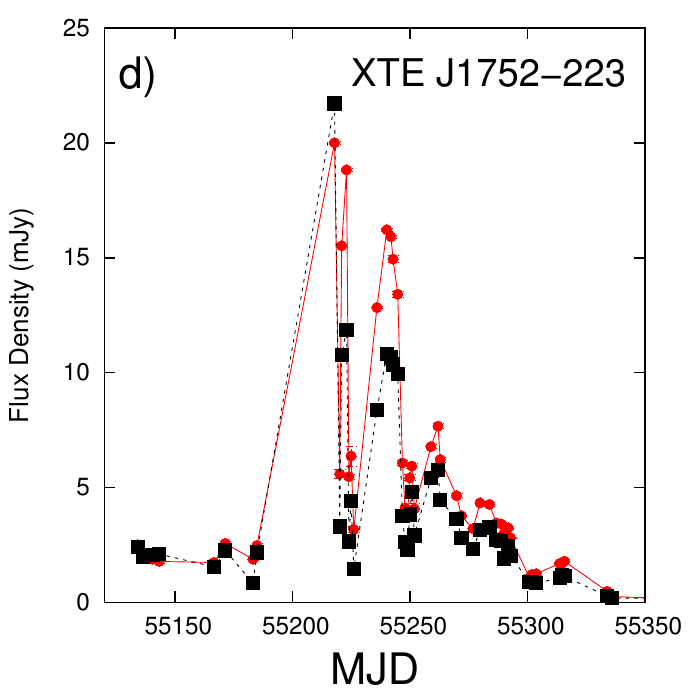}\\
\caption{The structured radio outbursts in \lsp, Cygnus X-3, and
  XTE\,1752-223: an outburst with flat/inverted spectrum is followed
  by optically thin events.  a) \lsi (GBI data, black squares:
  8.3\,GHz, red circles: 2.2\,GHz) \citep[see also Fig. 3
  in][]{massi15}. The black dashed line at MJD\,50888 indicates the
  optically thin flare of \lsi observed at 1.25\,mm wavelength by
  \citet{paredes00}, b-c) Cygnus X-3 (GBI data, black squares:
  8.3\,GHz, red circles: 2.2\,GHz). Here, quasi-periodic oscillations
  of the spectral index are shown in the inset, d) XTE J1752-223 (ATCA
  data, black squares: 9\,GHz, red circles: 5.5\,GHz, whereas errors
  are smaller than the symbols size).}
\label{other}
\end{figure}

During our Effelsberg campaign in 2012 March-April, \lsi was monitored
at seven frequencies between 2.6 and 32\,GHz over a total period of
24\,days with a cadence of about 12\,hours (over the first 12\,days). Our
observations reveal the following results:

\begin{enumerate}
\item The broad-band data show a main peak with nearly flat spectra
  followed by a sequence of small amplitude, optically thin events.
  This sequence of outbursts with different spectral characteristics
  is also found  in Cygnus X-3: Fig. \ref{other}b shows GBI
  observations of Cygnus X-3 \citep{waltman1995}, where a first
  outburst with almost flat spectrum ($\alpha \sim 0.05$) is seen,  followed by a minor optically thin flare.  Figure \ref{other}c
  displays a first outburst exhibiting an inverted spectrum ($\alpha
  \sim 0.6$) followed by a bursting phase of optically thin emission,
  as also seen for \lsi in the GBI observations of
  Fig. \ref{other}a. Clearly both sources, \lsi and Cygnus X-3, show
  the same sequence of events: a first outburst with flat/inverted
  spectrum is followed by one or more optically thin flares. A third
  microquasar with the same complex outburst behaviour is \xte{}
  \citep{brocksopp13}. In Fig. \ref{other}d we show one outburst at
  5.5/9\,GHz \citep[see also Fig. 1 in][]{brocksopp13}, where the
  first peak with $\alpha\sim 0.2$ is clearly distinguished from the
  subsequent optically thin flares.  \citet{brocksopp13} associate the
  first flare with the compact \xte{} jet and the sequence of
  optically thin flares with discrete ejections or shocks.

\item During the sequence of \lsi sub-flares presented in
  Fig. \ref{LCs}, $\alpha$ appears to oscillate in a quasi-regular
  fashion. Such spectral variability has so far only been discussed in the
  microquasar \grs \citep[see Fig. 2 in][]{fender02}.  As apparent
  from the inset of Fig. \ref{other}c (2.2/8.3\,GHz spectral index
  vs. time),  Cygnus X-3 also shows similar spectral index oscillations
  during the three, minor sub-flares occurring between MJD\,48470 and
  MJD\,48480.
\end{enumerate}

The dashed line in Fig. \ref{other}a indicates the epoch where
\citet{paredes00} obtained a steep \lsi cm/mm spectrum up to 250\,GHz
(1998 March 15) during one of the optically thin sub-flares following
the main flare. As a follow-up of our work, it is certainly of interest
to observe the main outburst of \lsi also at higher radio
frequencies. Indeed, flat/inverted spectra in microquasars have also been
established  at short millimeter wavelengths \citep{fender00,
  fender01} and it will be important to clarify if the short accretion
phase in the eccentric orbit of \lsi is able to maintain a flat
spectrum up to such high frequencies.

\acknowledgements This research is based on observations with the
100 m telescope of the MPIfR (Max-Planck-Institut f\"ur
Radioastronomie) at Effelsberg. We thank Fr\'ed\'eric Jaron and
J\"urgen Neidhofer for several interesting discussions and the
anonymous referee for useful comments.  The work is partly supported
by the German Excellence Initiative via the Bonn Cologne Graduate
School and the International Max Planck Research School for Astronomy
and Astrophysics.

\bibliographystyle{aa}

\begin{thebibliography}{}

\bibitem[Abdo et al.(2009)]{abdo09} Abdo, A.~A., et al.\ 2009, \apj, 701,
L123

\bibitem[Bondi \& Hoyle(1944)]{bondihoyle44} Bondi, H., \& Hoyle, F.\ 1944, \mnras, 104, 273

\bibitem[Bosch-Ramon et al.(2006)]{boschramon06} Bosch-Ramon, V., Paredes, J.~M., Romero, G.~E., 
\& Rib{\'o}, M.\ 2006, \aap, 459, L25

\bibitem[Brocksopp et al.(2013)]{brocksopp13} Brocksopp, C., 
        Corbel, S., Tzioumis, A., et al.\ 2013, \mnras, 432, 931 

\bibitem[Casares et al.(2005)]{casares05} Casares, J. et al.\ 2005, \mnras, 360, 1105

\bibitem[Chernyakova et al.(2014)]{chernyakova14} Chernyakova, M., 
Abdo, A.~A., Neronov, A., et al.\ 2014, \mnras, 439, 432 

\bibitem[Connors et al.(2002)]{connors02} Connors, T.~W. et al.\ 2002, \mnras, 336, 1201 

\bibitem[Corbel et al.(2013)]{corbel13} Corbel, S., Aussel, H., 
Broderick, J.~W., et al.\ 2013, \mnras, 431, L107 

\bibitem[Dhawan et al. (2006)]{dhawan06} Dhawan, V.,  Mioduszewski, A., \&
Rupen, M. 2006, Proceedings of  the VI Microquasar Workshop, p. 52.1

\bibitem[Dubus(2006)]{dubus06} Dubus, G. \ 2006, \aap, 456, 801

\bibitem[Fender et al.(2000)]{fender00} Fender, R.~P., Pooley, G.~G., Durouchoux, P., Tilanus, R.~P.~J., 
\& Brocksopp, C.\ 2000, \mnras, 312, 853 

\bibitem[Fender(2001)]{fender01} Fender, R.~P.\ 2001, \mnras, 
322, 31 

\bibitem[Fender et al.(2002)]{fender02} Fender, R.~P., Rayner, 
        D., Trushkin, S.~A., et al.\ 2002, \mnras, 330, 212 

\bibitem[Fuhrmann et al.(2008)]{fuhrmann08} Fuhrmann, L., Krichbaum, T.~P., Witzel, A., et al.\ 2008, \aap, 490, 1019 

\bibitem[Gregory et al.(1979)]{gregory79} Gregory, P.~C., Taylor,
A.~R., Crampton, D., et al.\ 1979, \aj, 84, 1030

\bibitem[Gregory(2002)]{gregory02} Gregory, P.~C.\ 2002, \apj, 575, 427

\bibitem[Jaron \& Massi(2013)]{jaronmassi13} Jaron, F., \& Massi, M.\ 2013, \aap, 559, AA129 

\bibitem[Jaron \& Massi(2014)]{jaronmassi14} Jaron, F., \& Massi, M.\ 2014, \aap, 572, AA105 

\bibitem[Kaiser(2006)]{kaiser06} Kaiser, C.~R.\ 2006, \mnras, 367, 1083

\bibitem[Maraschi \& Treves(1981)]{maraschitreves81} Maraschi, L., Treves, A.\ 1981, \mnras, 194, 1P

\bibitem[Marti \& Paredes(1995)]{martiparedes95} Marti, J., \& Paredes,
J.~M.\ 1995, \aap, 298, 151

\bibitem[Massi \& Kaufman Bernad{\'o}(2009)]{massikaufman09} Massi, M., \&
Kaufman Bernad{\'o}, M.\ 2009, \apj, 702, 1179

\bibitem[Massi \& Jaron(2013)]{massijaron13} Massi, M., \& Jaron, F.\ 2013, \aap, 554, A105

\bibitem[Massi \& Torricelli-Ciamponi(2014)]{massitorricelli14} Massi, M., \& Torricelli-Ciamponi, G.\ 2014, \aap, 564, A23 

\bibitem[Massi(2015)]{massi15} Massi, M.\ 2015, 
arXiv:1502.07543, PoS(EVN 2014)062

\bibitem[Mi{\v s}kovi{\v c}ov{\'a} et al.(2011)]{miskovica11} Mi{\v
    s}kovi{\v c}ov{\'a}, I., Hanke, M., Wilms, J., et al.\ 2011, Acta
  Polytechnica, 51, 85

\bibitem[Paredes et al.(2000)]{paredes00} Paredes, J.~M.,
  Mart{\'{\i}}, J., Peracaula, M., Pooley, G., \& Mirabel, I.~F.\
  2000, \aap, 357, 507

\bibitem[Romero et al.(2007)]{romero07} Romero, G.~E. et al. \ 2007, \aap, 474, 15

\bibitem[Strickman et al.(1998)]{1998ApJ...497..419S} Strickman, M.~S.,
Tavani, M., Coe, M.~J., et al.\ 1998, \apj, 497, 419

\bibitem[Taylor et al.(1992)]{taylor92} Taylor, A.~R., Kenny, 
H.~T., Spencer, R.~E., \& Tzioumis, A.\ 1992, \apj, 395, 268 

\bibitem[Waltman et al.(1995)]{waltman1995} Waltman, E.~B., 
Ghigo, F.~D., Johnston, K.~J., et al.\ 1995, \aj, 110, 290

\bibitem[Zimmermann(2013)]{zimmermann13}
Zimmermann, L.: Variability of radio and TeV emitting X-ray binary systems
 The case of \lsp. Bonn, Univ., Diss., 2013 URN:urn:nbn:de:hbz:5n-33175

\end{thebibliography}

\end{document}